\documentclass[reprint, amsmath, amssymb, aps, prl]{revtex4-2}

\usepackage{graphicx}
\usepackage{dcolumn}
\usepackage{bm}
\usepackage{hyperref}
\usepackage{physics}
\usepackage{xfrac}
\usepackage{xcolor}
\usepackage{soul}

\newcommand{\comment}[1]{}

\begin{document}

\preprint{APS/123-QED}

\title{Observation of coherent coupling between super- and subradiant states of an ensemble of cold atoms collectively coupled to a single propagating optical mode}

\author{Riccardo Pennetta}
\email{riccardo.pennetta@hu-berlin.de}
\author{Daniel Lechner}
\author{Martin Blaha}
\author{Arno Rauschenbeutel}
\author{Philipp Schneeweiss}
\author{J\"urgen Volz}
\email{juergen.volz@hu-berlin.de}

\affiliation{%
 Department of Physics, Humboldt Universit\"at zu Berlin, 12489 Berlin, Germany\\
}%

\date{\today}

\begin{abstract}
We discuss the evolution of the quantum state of an ensemble of atoms that are coupled via a single propagating optical mode. We theoretically show that the quantum state of $N$ atoms, which are initially prepared in the timed Dicke state, evolves through all the $N-1$ states that are subradiant with respect to the propagating mode.
We predict this process to occur for any atom number and any atom--light coupling strength. These findings are supported by measurements performed with cold cesium atoms coupled to the evanescent field of an optical nanofiber. We experimentally observe the evolution of the state of the ensemble passing through the first two subradiant states, leading to sudden, temporary switch-offs of the optical power emitted into the nanofiber. Our results contribute to the fundamental understanding of collective atom--light interaction and apply to all physical systems, whose description involves timed Dicke states.

\end{abstract}

\maketitle


The collective interaction of $N$ quantum emitters with light can differ substantially depending on whether the distance between any two of the emitters is significantly smaller or larger than the wavelength of the light field.
The former situation has been initially investigated by Dicke \cite{Dicke1954}, who decomposed the Hilbert space of the system in superradiant states, whose decay rate towards lower energy states is enhanced with respect to the one of an isolated emitter, and subradiant states, for which decay to the ground state is suppressed. In particular, in the single excitation limit which we consider throughout this Letter, the Dicke model identifies a single superradiant state with an $N$ times enhanced decay rate, while there are $N-1$ subradiant states.

\begin{figure}[]
\includegraphics[width=0.95\linewidth]{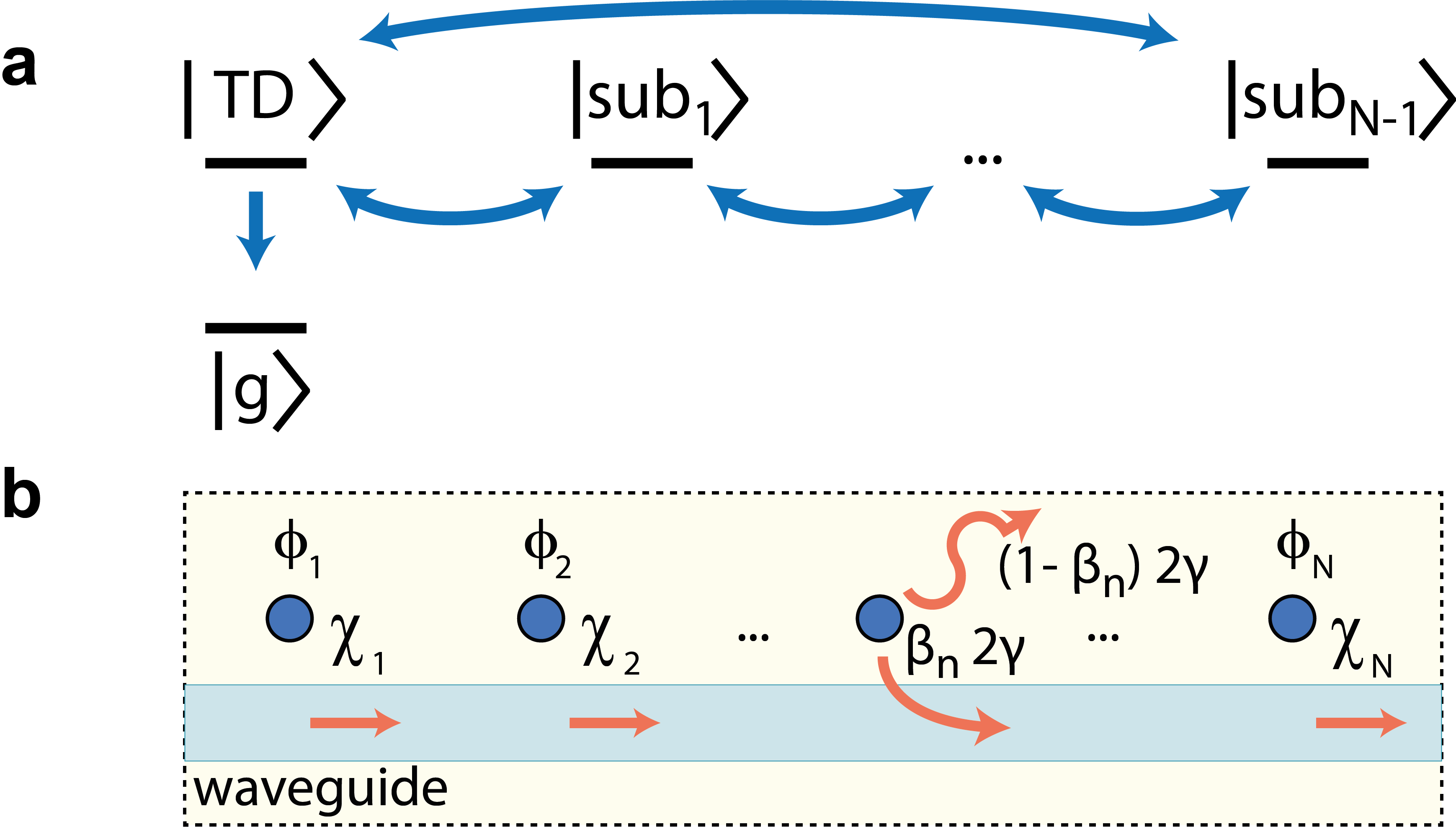}
\caption{\label{fig:PhysicalSetting} (a) Collective states for an ensemble of $N$ atoms in the single excitation limit. TD: timed Dicke state, sub$_n$: $n^\textrm{th}$ subradiant state, g: ground state. (b) $N$ atoms coupled to a single-mode optical waveguide. The factors $\beta_n \, 2\gamma$ and $ (1-\beta_n) \, 2\gamma$ represent the photon emission rate of the $n^\textrm{th}$ atom into guided and unguided optical modes, respectively. The excited state amplitude of the $n^\textrm{th}$ atom is indicated by $\phi_n$, while $\chi_n$ is the light field amplitude right after the $n^\textrm{th}$ atom. 
}
\end{figure}

A formal description of the time evolution of extended ensembles of emitters 
(i.e., inter-atomic distance between any two emitters $> \lambda$) 
within the same framework of the Dicke model requires the introduction of the so-called timed Dicke states \cite{Scully2006}, which can be prepared, e.g., by exciting the emitters with a single photon with wavevector $\boldsymbol{k}$:
\begin{eqnarray}
\ket{\textrm{TD}} = \dfrac{1}{\sqrt{N}} \sum_{n=1}^{N}  e^{i \boldsymbol{k} \cdot \boldsymbol{r}_\textbf{\textit{n}}} \hat{\sigma}^+_n \ket{0} ~.
\label{eq:TimedDickeState}
\end{eqnarray}
Here, $\boldsymbol{r}_\textbf{\textit{n}}$ and $\hat{\sigma}^+_n$ indicate the position and the raising operator for the $n^\textrm{th}$ emitter in the ensemble and $\ket{0}=\ket{0_\textrm{ph}}\otimes \ket{g_1, ..., g_N}$, where $\ket{0_\textrm{ph}}$ is the vacuum state and $\ket{g_1, ..., g_N}$ indicates that all atoms are in the ground state.  An ensemble in the timed Dicke state experiences superradiant decay and collectively enhanced emission of light into the same optical mode that excited the system. Likewise, it is possible to introduce $N-1$ states, that are orthogonal to the timed Dicke state and subradiant with respect to the optical mode with wavevector $\boldsymbol{k}$.
This approach is suitable for describing a considerable number of physical systems, including for instance cold atom clouds \cite{Araujo2016, Roof2016, Guerin2016, Ferioli2021, Goban2015, Okaba2019, Bettles2020, Pennetta2021}, Rydberg atoms \cite{ParisMandoki2017, Stiesdal2020} and emitters in solid state samples \cite{Rohlsberger2010}.
In addition, in cavity quantum electrodynamics the timed Dicke physics is at the basis of the assumptions of the Tavis-Cummings model \cite{Tavis1968, Blaha2021}, which describes collectively enhanced light-matter coupling between an atomic ensemble and a single mode cavity \cite{Colombe2007, Brennecke2007}.
For these reasons, notable effort has been recently devoted to describe the time dynamics of extended ensembles \cite{Scully2006, Svidzinsky2008, Kumlin2020, Jen2020, Berman2020, Pivovarov2021}, a task which, unlike the standard Dicke model, needs to consider coupling between super- and subradiant states, see Fig. \ref{fig:PhysicalSetting}(a).

In this Letter, we study 
a one-dimensional ensemble of atoms coupled via a single optical mode, and we discuss the time evolution of its state.
We present a theoretical approach that unveils the microscopic 
temporal dynamics of the system within a simple mathematical framework. We show that during its temporal evolution, an ensemble initially prepared in the timed Dicke state evolves through all basis states of the Hilbert space, reaching, one after the other, all the $N-1$ subradiant states.
This process occurs for any atom number (unlike the case of three-dimensional atom clouds \cite{Svidzinsky2008}) and for any coupling strength. Notably, it already appears for two emitters with arbitrarily weak coupling to the optical mode \cite{Kien2017}.
To support the predictions of our model, we present experimental results that we obtain interfacing an ensemble of cold cesium (Cs) atoms with the evanescent field part of the guided-mode of an optical nanofiber. After exciting the atoms with boxcar-shaped nanofiber-guided laser pulses with a switch-off time much shorter than the atomic lifetime, we investigate the temporal response of the system by recording the optical power exiting the nanofiber.
We observe the passage of the state of the ensemble through the first two subradiant states with respect to the guided mode, which leads to abrupt and temporary drops in the nanofiber-guided light field.



A sketch of the considered configuration is shown in Fig.~\ref{fig:PhysicalSetting}(b).
An ensemble of $N$ atoms 
is coupled to a single guided optical mode. The atom-waveguide coupling strength is characterized by the parameter $\beta_{n}$, defined as the ratio between the spontaneous emission rate of the $n^\textrm{th}$ atom into the considered waveguide mode and the total photon emission rate of a single-atom into all modes, $2 \gamma$, see Fig.~\ref{fig:PhysicalSetting}(b). Throughout this Letter, we assume unidirectional propagation, an approximation which is justified by the collective enhancement of forward emission typical of the timed Dicke state \cite{Scully2006, Pennetta2021} and which holds as long as the atoms are not arranged at the Bragg condition (i.e., inter-atomic distance equal to an integer multiple of $\frac{\lambda}{2}$). In addition, for simplicity we also assume that all atoms have the same coupling strength $\beta$ to the waveguide, a situation which describes well, e.g., trapped atoms \cite{Vetsch2010, Thompson2013, Goban2015}. However, our model can be easily extended to the general case, as shown in the Supplemental Material.

The general quantum state of the system can be written as:
\begin{align}
\nonumber
\ket{\psi(t)} =& \bigg[ \sum_{n=1}^N e^{ik z_n}\phi_n(t)\hat{\sigma}^+_n +
\\
&+ \int_{-\infty}^{\infty} e^{ik z_n} \chi(t,z) \hat{a}_z^{\dagger} dz \bigg] \ket{0} +
\ket{\psi_{fs}(t)} ~,
\label{eq:GeneralState}
\end{align}
where $\phi_n(t)$ and $z_n$ represent the excited state amplitude and the position along the fiber of the $n^\textrm{th}$ atom in the ensemble, respectively. In addition, $\chi(t,z)$ indicates the complex amplitude of the waveguide-coupled light field, $\hat{a}_z^{\dagger}$ is the creation operator for a forward-propagating photon at position $z$, and $\ket{\psi_{fs}(t)}$ denotes the state of the photons that are emitted into free space.

The collective decay rate of the ensemble, $\Gamma_{ens}(t)$, is defined as the ratio between the energy loss rate and the energy stored in the ensemble:
\begin{eqnarray}
\Gamma_{ens}(t) = - \frac{\sum_{n=1}^{N} \tfrac{\partial}{\partial t} \lvert\phi_n(t) \rvert^2 }{\sum_{n=1}^{N} \lvert\phi_n(t) \rvert^2 } = \Gamma_{fs} + \Gamma_{ens, wg}(t) ~,
\label{eq:Gamma_coll}
\end{eqnarray}
where $\Gamma_{fs}=(1-\beta)2\gamma$ is the decay rate into free space, that we assume to be the same for all atoms. The second term, $\Gamma_{ens, wg}(t)$, indicates the ensemble decay rate into the waveguide, defined as the energy leaving the ensemble via guided light divided by the stored energy:
\begin{eqnarray}
\Gamma_{ens, wg}(t) = 
\frac{ v_g \lvert \chi_N(t) \rvert^2 }{\sum_{n=1}^{N} \lvert\phi_n(t) \rvert^2 } ~,
\label{eq:Gamma_coll_wg}
\end{eqnarray}
where $\chi_{N}(t)$ is the complex amplitude of the light field right after the $N^\textrm{th}$ atom and  $v_g$ is the group velocity of the guided optical mode.
We note that  $\Gamma_{ens}(t)$ is sometimes inferred from the decay rate of the light intensity emitted by the atoms into the considered mode, $\Gamma_{light}(t)$, \cite{Guerin2016, Bettles2020, Roof2016, Stiesdal2020, Pennetta2021}, i.e.:
\begin{equation}
\Gamma_{light}(t)=-\dfrac{\partial}{\partial t} \left[ \textrm{log} \left( \dfrac{|\chi_N(t)|^2}{|\chi_N(0)|^2} \right) \right] = - \frac{ \tfrac{\partial}{\partial t} \lvert \sum_{n=1}^{N} \phi_n(t) \rvert^2 }{\sum_{n=1}^{N} \lvert\phi_n(t) \rvert^2 } ~.
\label{eq:Gamma_light}
\end{equation}
However, while convenient to access experimentally, a comparison between Eqs.~(\ref{eq:Gamma_coll}) and (\ref{eq:Gamma_light}) shows that this quantity is in general not related to $\Gamma_{ens} (t)$, whose estimation requires access to the complete state of the system. Interestingly, $\Gamma_{ens}$ and $\Gamma_{light}$ coincide for the timed Dicke state.

To theoretically describe the time-evolution of $\ket{\psi(t)}$, we follow the approach of \cite{Shen2009, Blaha2021, Pennetta2021}, which, starting from a real-space Hamiltonian, allows us to calculate the excitation amplitudes, $\phi_n$, in the steady state by solving the time-independent Schr\"odinger equation. In the low excitation limit, the dynamics of the system in the time-domain can be then obtained via Fourier analysis (see Supplemental Material).

For a system that is prepared in the timed Dicke state, i.e., $\phi_n(t=0)=1/\sqrt{N}$, with $n=1,...,N$, and that then evolves freely, our model allows us to derive analytic expressions for $\phi_n(t)$ and $\chi_{n}(t)$, where the latter is proportional to the sum of the excited state amplitudes of the first $n$ atoms:
\begin{equation}
\phi_n(t) = \dfrac{1}{\sqrt{N}} e^{-\gamma t}L_{n-1}^{(0)}(2 \beta \gamma t) ~,
\label{eq:SolutionsTimedDickeStateAtoms}
\end{equation}
\begin{equation}
\chi_n(t) = \dfrac{\sqrt{2 \beta \gamma}}{i \sqrt{v_g}} \sum_{m=1}^n \phi_m(t) 
= \dfrac{\sqrt{2 \beta \gamma}}{i \sqrt{v_g N}}e^{-\gamma t}L_{n-1}^{(1)} (2 \beta \gamma t) ~,
\label{eq:SolutionsTimedDickeStateLight}
\end{equation}
where $L_{m}^{(\alpha)}$ are the generalized Laguerre polynomials.
The solutions show that, with the exception of the factor $e^{-\gamma t}$, the entire dynamics only depends on the product $\beta \gamma t$ and therefore it is essentially the same for any emitter-mode coupling strength, $\beta$, after appropriate scaling of the observation time. In the limit $\beta = 1$, these results agree with calculations based on a master equation approach \cite{Kumlin2020}.


\begin{figure}[]
\includegraphics[width=0.95\linewidth]{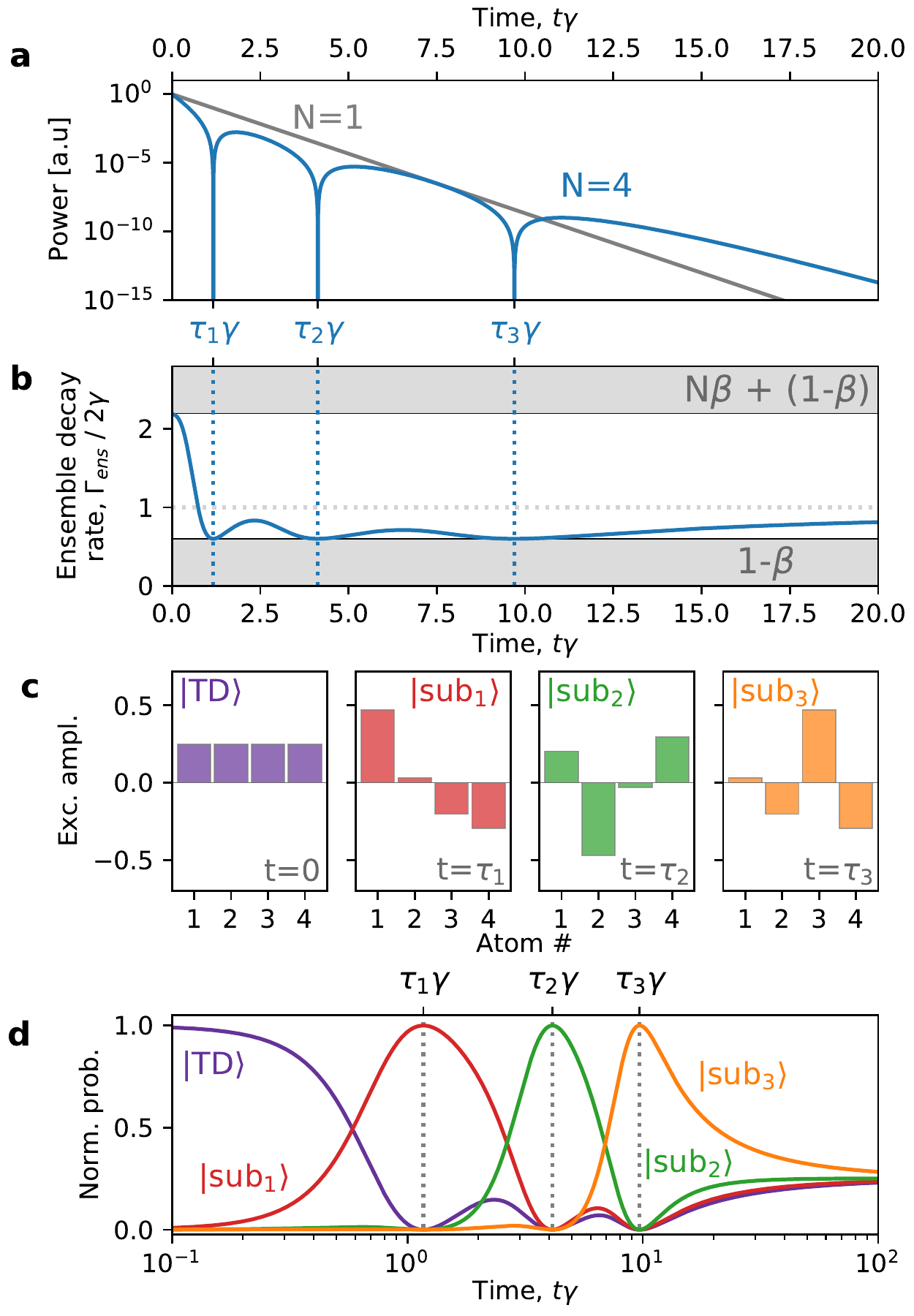}
\caption{\label{fig:Theory}
(a) Calculated waveguide-coupled power, $|\chi(t)|^2$, (normalized at $t=0$) as a function of time for $N=4$ atoms (blue line) with $\beta=0.4$. For comparison, the case for $N=1$ atom is shown as a gray line.
(b) Ensemble decay rate, $\Gamma_{ens}(t)$, as a function of normalized time. The vertical blue dashed lines indicate the instants corresponding to which the ensemble is in a subradiant state with respect to the waveguide, $t = \tau_i$, $i = 1, 2, 3$.
(c) Normalized excited state amplitude for the atoms in the ensemble for the timed Dicke state and the subradiant states.
(d) Projection on the timed Dicke state and the subradiant states of the residual excitation stored in the ensemble as a function of time.
}
\end{figure}

Equations~(\ref{eq:SolutionsTimedDickeStateAtoms}) and~(\ref{eq:SolutionsTimedDickeStateLight}) completely determine the state of the system and allow one to calculate quantities of interest such as the total waveguide-coupled optical power, $|\chi(t)|^2$, and $\Gamma_{ens}(t)$, which, as an example, are shown in Fig.~\ref{fig:Theory}(a) and (b), respectively, for the case $N=4$ and $\beta=0.4$.
At $t=0$ the ensemble is in the timed Dicke state and exhibits a superradiant decay rate into the waveguide mode, which is $N$ times enhanced with respect to the one of a single atom: $\Gamma_{ens,wg}(0) = N \cdot \beta \, 2\gamma$.
For $t>0$, unlike for the predictions of the standard Dicke model, the waveguide-coupled light power exhibits $N-1$  zeros, whose times of occurrence, $\tau_m$ ($m=1, ..., N-1$), are the roots of $L_{N-1}^{(1)} (2 \beta \gamma t)$. This relation explicitly defines a set of states, which are subradiant with respect to the waveguide:
\begin{eqnarray}
\ket{\textrm{sub}_\textrm{m}} = \sum_{n=1}^N e^{ik z_n}\phi_n(\tau_m)\hat{\sigma}^+_n \ket{0} ~.
\label{eq:SubradiantStates}
\end{eqnarray}
For these states, we have $\Gamma_{ens,wg}(\tau_m) = 0$. However, the emission into free space is unmodified, therefore $\Gamma_{ens} (\tau_m) = \Gamma_{fs}$, which is also the minimum total decay rate that can be obtained in the considered configuration, see Fig.~\ref{fig:Theory}(b).
 
Figure~\ref{fig:Theory}(c) shows the excited state amplitudes for the super- and subradiant states. For instance, for the first subradiant state, $\ket{ \textrm{sub}_1 }$, the first and last two atoms of the ensemble radiate in phase opposition, causing fully destructive interference of the waveguide-coupled light fields.
The subsequent subradiant states are characterized by a similar structure but the number of sub-ensembles that radiate with opposite phases increases progressively by one.
These states are all orthogonal to each other and to the timed Dicke state (i.e., the superradiant state). All together, these states form a complete basis of the subspace in which a single excitation is shared by the atoms, allowing a full description of the state of the ensemble (see Supplemental Material).

To shed further light on the system dynamics, Fig.~\ref{fig:Theory}(d) shows the decomposition onto this basis of the residual excitation stored in the atoms as a function of time. The ensemble, initially prepared in the timed Dicke state, passes through all the subradiant states and eventually remains (for $t \gg \tau_{N-1}$) in a superposition of the timed Dicke- and the subradiant states with equal probability of $1/N$. At this point, $\Gamma_{ens}(t)$ asymptotically converges to the single-atom decay rate, i.e., $2 \gamma$, see Fig.~\ref{fig:Theory}(b). This process remains qualitatively unaltered for any atom number $N$ and coupling strength $\beta$.


It is interesting to note that for $\beta \ll 1$ and large $N$ the time evolution of the ensemble becomes independent of the individual atom-mode coupling strength and only depends on the optical depth, OD $\simeq 4 \beta N$. Indeed, under these conditions, we can approximate Eqs.~(\ref{eq:SolutionsTimedDickeStateAtoms}) and~(\ref{eq:SolutionsTimedDickeStateLight}) as:
\begin{eqnarray}
\phi_n(t) \simeq & \dfrac{1}{\sqrt{N}} e^{-\gamma t} J_{0}\bigg( \sqrt{2 \gamma\, \textrm{OD}\, \frac{n-1}{N} t} \bigg) ~, \qquad
\label{eq:SolutionsTimedDickeStateAtomsBessel}
\end{eqnarray}
\begin{eqnarray}
\chi_N(t) \simeq & \dfrac{\sqrt{\textrm{OD}}}{i \sqrt{4v_g \beta t}}e^{-\gamma t} J_{1} (\sqrt{2 \gamma \, \textrm{OD} \, t}) ~, \quad
\label{eq:SolutionsTimedDickeStateLightBessel}
\end{eqnarray}
where $J_\alpha(x)$ are the Bessel functions of the first kind (see Supplemental Material). These relations are relevant for experiments involving large numbers of emitters and agree with results obtained for continuous resonant media \cite{Buerck1999}. The fact that the argument of the Bessel functions only depends on the OD suggests that the dynamics of the system is also well described by Eqs.~(\ref{eq:SolutionsTimedDickeStateAtomsBessel}) and (\ref{eq:SolutionsTimedDickeStateLightBessel}) when the coupling strength varies from atom to atom as long as $\beta_n\ll 1$, see Fig.~\ref{fig:PhysicalSetting}. We numerically confirmed that this is indeed the case.



\begin{figure}[]
\includegraphics[width=.9\linewidth]{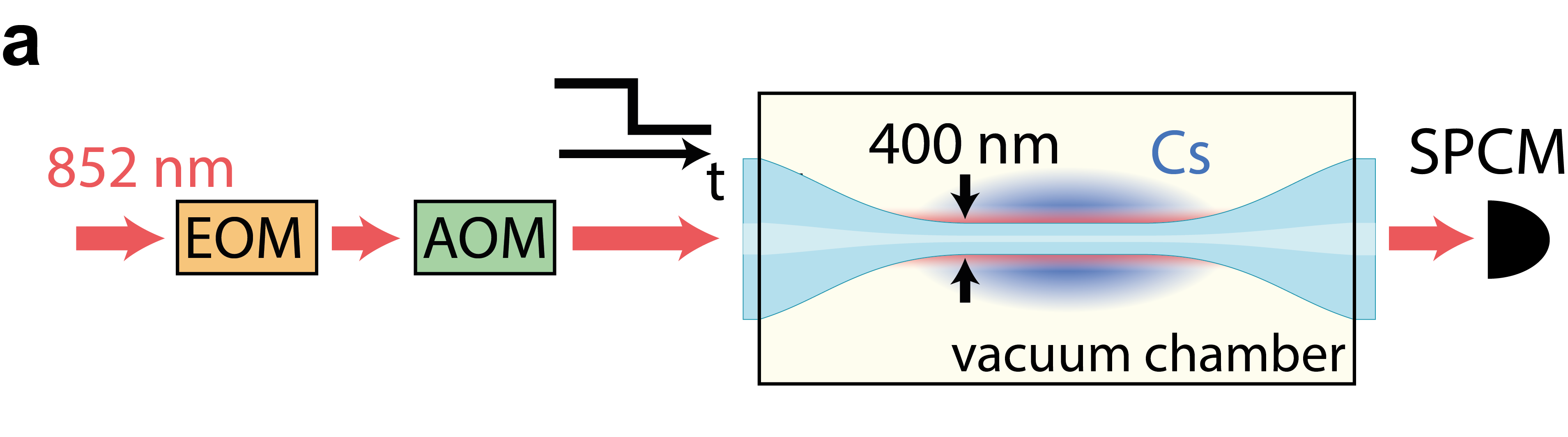}
\includegraphics[width=.94\linewidth]{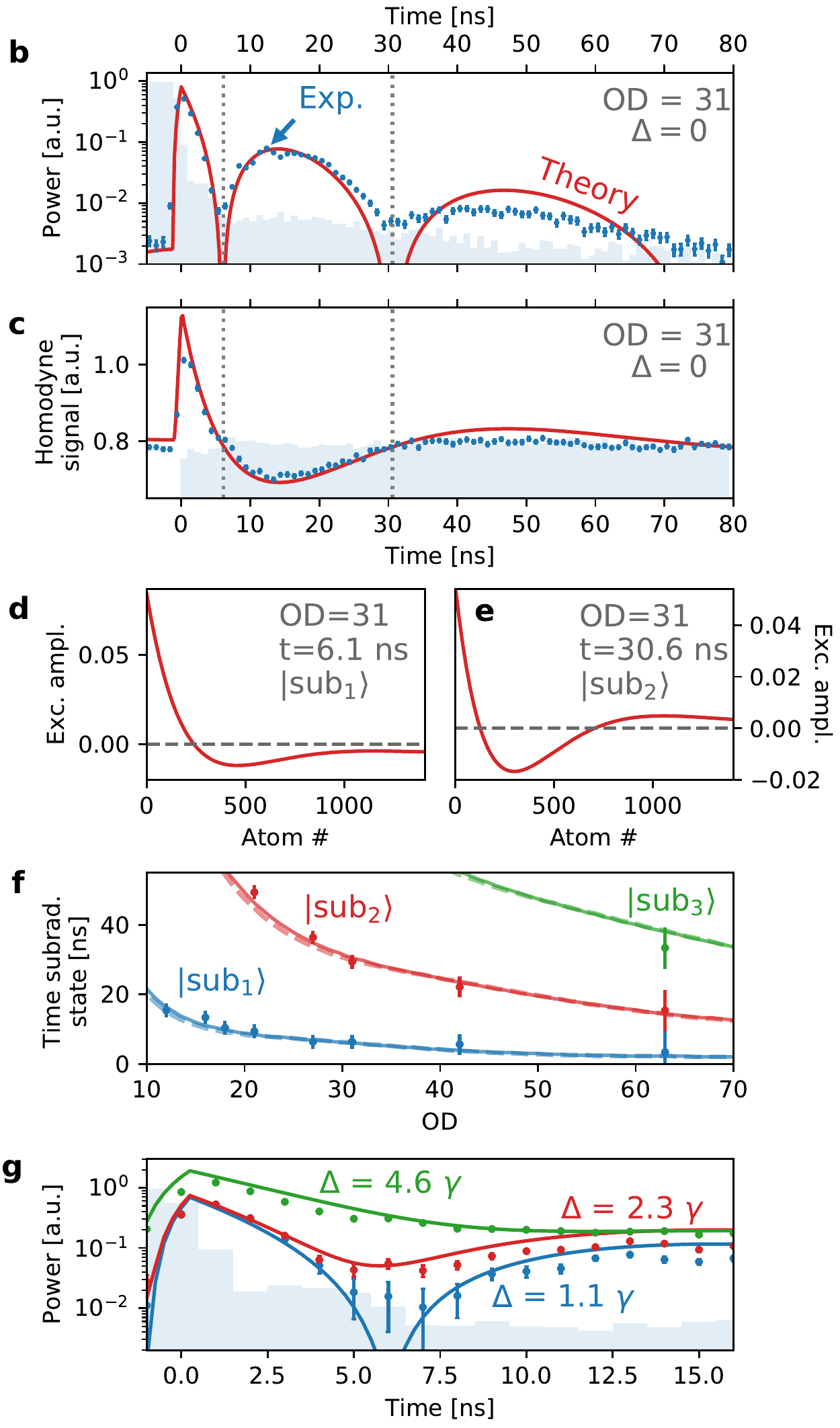}
\caption{\label{fig:Exp_data} (a) Experimental setup. Cs: cesium, EOM: electro-optic modulator, AOM: acousto-optic modulator, SPCM: single photon counting module. (b) Measured (blue dots) and predicted (red line) optical power exiting the nanofiber for $\Delta = 0$ and OD $=31$. The blue shaded area shows the measured pulse in the absence of atoms. (c) Measured (blue dots) and predicted (red line) homodyne signal between the atoms-emitted light and a local oscillator for the same parameters as in (b). (d, e) Calculated excited state amplitudes of each atom for the parameters as in (b) for the first (d) and second (e) subradiant state. These states are reached at $t=6.1$~ns and $t=30.6$~ns, as indicated in (b, c) with dashed gray lines. (f) Time at which the system reaches the first (blue line and dots), second (red line and dots) and third (green line and dot) subradiant states as a function of the OD. The measurements for OD of 42 and 63 have been obtained with multiple passes of the pulse through the ensemble, see Suppl. Mat. The dots are measured data, while the solid and dashed lines are theoretical predictions for $\beta = 0.55\%$ and $\beta = 20\%$, respectively. (g) Measured optical power for OD $=28$ and $\Delta$ = 1.1 $\gamma$ (blue dots), $\Delta$ = 2.3 $\gamma$ (red dots) and $\Delta$ = 4.6 $\gamma$ (green dots). The solid lines are theoretical predictions.
}
\end{figure}



In the one-dimensional geometry considered here, the timed Dicke state can be approximately prepared by exciting the atoms with very short laser pulses, carrying an average energy corresponding to less than one photon. However, this method is very inefficient because most of the time the excitation pulse passes through the ensemble without interacting with the atoms, complicating the observation of the atom-emitted light with good signal to noise ratio.
To experimentally explore the concepts discussed above, we therefore preferred to employ long boxcar-shaped excitation pulses. They prepare a state, whose time dynamics has the same features of the one of the timed Dicke state but can be prepared with significantly higher probability, see Supplemental Material.



Fig.~\ref{fig:Exp_data}(a) shows a sketch of the experimental setup. A single-mode optical nanofiber with a 400-nm diameter and a waist length of 1 cm is placed in a vacuum chamber and surrounded by a cloud of cold Cs atoms from a magneto-optical trap (MOT). Nanofiber-guided excitation pulses that are near-resonant with the Cs D2 transition ($6S_{1/2}$, $F=4 \rightarrow 6P_{3/2}, F^\prime=5$) are generated from an incident cw laser beam using an electro-optic modulator followed by an acousto-optic modulator. The $<1$-ns fall-time (90$\%$-10$\%$) of the pulses is much shorter than the lifetime of the excited state ($\tau$ = $1/2\gamma$ = 30.4 ns, $\gamma/2\pi$ = 2.6 MHz~[\onlinecite{Steck2019}]), while the total pulse length (100~ns $>$ 3$\tau$) is long enough for the system to approximately reach a steady state.
The optical power transmitted through the nanofiber is measured using a single-photon counting module. We note that each excitation pulse has a mean power significantly smaller than one single photon energy per atomic lifetime. This ensures that, on average, less than one excitation is stored in the atomic ensemble.


Time-resolved measurements of the emitted optical power are shown in Fig.~\ref{fig:Exp_data}(b) for probe pulses resonant with the atomic transition, i.e., detuning $\Delta=\omega - \omega_a=0$, where $\omega_a=0$ is the atomic resonance frequency, and for OD $=31$.
Right after the switch-off of the excitation pulse at $t=0$, the ensemble initially decays at a superradiant rate, reaching minima in transmission around $t = 6.1$~ns and $t = 30.6$~ns. Fig.~\ref{fig:Exp_data}(d, e) depicts the calculated excited state amplitudes for the atoms in the ensemble at these moments in time. Complete destructive interference of the light emitted from atoms into the guided mode is predicted, i.e., the ensemble is in a subradiant state with respect to the nanofiber-guided mode.

Our theoretical model predicts a change in the sign of the projection of the ensemble state on the timed Dicke state and thus of the emitted light field amplitude each time the system passes through one of the subradiant states. To experimentally investigate this feature, we repeat the measurement of Fig.~\ref{fig:Exp_data}(b), this time interfering the light emitted by the atomic ensemble with a local oscillator. The result is shown in Fig.~\ref{fig:Exp_data}(c), in which clear constructive and destructive interference with the local oscillator is observed before and after the first subradiant state respectively, in very good agreement with our predictions.

Fig.~\ref{fig:Exp_data}(f) illustrates the time at which the system is in the first (blue points and lines), second (red points and lines) and third (green point and lines) subradiant state as a function of the OD. The theoretical predictions shown as solid lines have been calculated for $\beta=0.55\%$, which is the average value measured in our experiment~[\onlinecite{Johnson2019}], while the dashed lines correspond to a much larger value of $\beta=20\%$ (approximately the highest $\beta$ that can be achieved in a nanofiber atom-interface \cite{Kien2005}). The close agreement between the two theory predictions and the experimental data confirms that for $N \gg 1$ the time evolution of the system depends on OD rather than on atom number.


Excitation via moderately detuned laser pulses results in an additional phase shift between subsequent atoms due to the dispersive response of the ensemble, which prevents complete destructive interference of the collective emission.
This was experimentally investigated by launching into the nanofiber pulses, whose central frequency is detuned with respect to the atomic transition. The results are shown in Fig.~\ref{fig:Exp_data}(g), in which the transmitted power for a time interval close to when the system reaches the first subradiant state is depicted for several detunings $\Delta$. 
As expected, compared to the case $\Delta=0$, larger detunings cause a gradual disappearance of the first minimum in the transmitted optical power.




In conclusion, we have shown that an ensemble of $N$ emitters prepared in the timed Dicke state with respect to a given propagating mode evolves through $N-1$ subradiant states.
This results in aperiodic switch-offs of the collectively emitted optical power. This non-monotonous decay originates from the dynamics induced by atom--atom coupling via propagating photons.
Our predictions are in excellent agreement with power and phase-sensitive measurements of the light emitted by an ensemble of Cs atoms coupled to the guided mode of an optical nanofiber.
Beyond providing fundamental insight into collective atom--light coupling, our results are of direct relevance for systems whose dynamics involve timed Dicke states as, e.g., optical quantum memories \cite{Lvovsky2009, Simon2010} and atomic ensembles in cavity and waveguide quantum electrodynamics \cite{Mivehvar2021, Johnson2019, Goban2015}.


\begin{acknowledgments}
We acknowledge financial support by the Alexander von Humboldt Foundation in the framework of an Alexander von Humboldt Professorship endowed by the Federal Ministry of Education and Research and by the Austrian Science Fund (NanoFiRe grant project No.
P31115).
\end{acknowledgments}

\bibliography{main.bib}

\providecommand{\noopsort}[1]{}\providecommand{\singleletter}[1]{#1}%
\begin{thebibliography}{33}%
\makeatletter
\providecommand \@ifxundefined [1]{%
 \@ifx{#1\undefined}
}%
\providecommand \@ifnum [1]{%
 \ifnum #1\expandafter \@firstoftwo
 \else \expandafter \@secondoftwo
 \fi
}%
\providecommand \@ifx [1]{%
 \ifx #1\expandafter \@firstoftwo
 \else \expandafter \@secondoftwo
 \fi
}%
\providecommand \natexlab [1]{#1}%
\providecommand \enquote  [1]{``#1''}%
\providecommand \bibnamefont  [1]{#1}%
\providecommand \bibfnamefont [1]{#1}%
\providecommand \citenamefont [1]{#1}%
\providecommand \href@noop [0]{\@secondoftwo}%
\providecommand \href [0]{\begingroup \@sanitize@url \@href}%
\providecommand \@href[1]{\@@startlink{#1}\@@href}%
\providecommand \@@href[1]{\endgroup#1\@@endlink}%
\providecommand \@sanitize@url [0]{\catcode `\\12\catcode `\$12\catcode
  `\&12\catcode `\#12\catcode `\^12\catcode `\_12\catcode `\%12\relax}%
\providecommand \@@startlink[1]{}%
\providecommand \@@endlink[0]{}%
\providecommand \url  [0]{\begingroup\@sanitize@url \@url }%
\providecommand \@url [1]{\endgroup\@href {#1}{\urlprefix }}%
\providecommand \urlprefix  [0]{URL }%
\providecommand \Eprint [0]{\href }%
\providecommand \doibase [0]{https://doi.org/}%
\providecommand \selectlanguage [0]{\@gobble}%
\providecommand \bibinfo  [0]{\@secondoftwo}%
\providecommand \bibfield  [0]{\@secondoftwo}%
\providecommand \translation [1]{[#1]}%
\providecommand \BibitemOpen [0]{}%
\providecommand \bibitemStop [0]{}%
\providecommand \bibitemNoStop [0]{.\EOS\space}%
\providecommand \EOS [0]{\spacefactor3000\relax}%
\providecommand \BibitemShut  [1]{\csname bibitem#1\endcsname}%
\let\auto@bib@innerbib\@empty
\bibitem [{\citenamefont {Dicke}(1954)}]{Dicke1954}%
  \BibitemOpen
  \bibfield  {author} {\bibinfo {author} {\bibfnamefont {R.~H.}\ \bibnamefont
  {Dicke}},\ }\bibfield  {title} {\bibinfo {title} {Coherence in spontaneous
  radiation processes},\ }\href {https://doi.org/10.1103/physrev.93.99}
  {\bibfield  {journal} {\bibinfo  {journal} {Physical Review}\ }\textbf
  {\bibinfo {volume} {93}},\ \bibinfo {pages} {99} (\bibinfo {year}
  {1954})}\BibitemShut {NoStop}%
\bibitem [{\citenamefont {Scully}\ \emph {et~al.}(2006)\citenamefont {Scully},
  \citenamefont {Fry}, \citenamefont {Ooi},\ and\ \citenamefont
  {W{\'{o}}dkiewicz}}]{Scully2006}%
  \BibitemOpen
  \bibfield  {author} {\bibinfo {author} {\bibfnamefont {M.~O.}\ \bibnamefont
  {Scully}}, \bibinfo {author} {\bibfnamefont {E.~S.}\ \bibnamefont {Fry}},
  \bibinfo {author} {\bibfnamefont {C.~H.~R.}\ \bibnamefont {Ooi}},\ and\
  \bibinfo {author} {\bibfnamefont {K.}~\bibnamefont {W{\'{o}}dkiewicz}},\
  }\bibfield  {title} {\bibinfo {title} {Directed spontaneous emission from an
  extended ensemble of {N} atoms: Timing is everything},\ }\bibfield  {journal}
  {\bibinfo  {journal} {Physical Review Letters}\ }\textbf {\bibinfo {volume}
  {96}},\ \href {https://doi.org/10.1103/physrevlett.96.010501}
  {10.1103/physrevlett.96.010501} (\bibinfo {year} {2006})\BibitemShut
  {NoStop}%
\bibitem [{\citenamefont {Ara{\'{u}}jo}\ \emph {et~al.}(2016)\citenamefont
  {Ara{\'{u}}jo}, \citenamefont {Kre{\v{s}}i{\'{c}}}, \citenamefont {Kaiser},\
  and\ \citenamefont {Guerin}}]{Araujo2016}%
  \BibitemOpen
  \bibfield  {author} {\bibinfo {author} {\bibfnamefont {M.~O.}\ \bibnamefont
  {Ara{\'{u}}jo}}, \bibinfo {author} {\bibfnamefont {I.}~\bibnamefont
  {Kre{\v{s}}i{\'{c}}}}, \bibinfo {author} {\bibfnamefont {R.}~\bibnamefont
  {Kaiser}},\ and\ \bibinfo {author} {\bibfnamefont {W.}~\bibnamefont
  {Guerin}},\ }\bibfield  {title} {\bibinfo {title} {Superradiance in a large
  and dilute cloud of cold atoms in the linear-optics regime},\ }\bibfield
  {journal} {\bibinfo  {journal} {Physical Review Letters}\ }\textbf {\bibinfo
  {volume} {117}},\ \href {https://doi.org/10.1103/physrevlett.117.073002}
  {10.1103/physrevlett.117.073002} (\bibinfo {year} {2016})\BibitemShut
  {NoStop}%
\bibitem [{\citenamefont {Roof}\ \emph {et~al.}(2016)\citenamefont {Roof},
  \citenamefont {Kemp}, \citenamefont {Havey},\ and\ \citenamefont
  {Sokolov}}]{Roof2016}%
  \BibitemOpen
  \bibfield  {author} {\bibinfo {author} {\bibfnamefont {S.}~\bibnamefont
  {Roof}}, \bibinfo {author} {\bibfnamefont {K.}~\bibnamefont {Kemp}}, \bibinfo
  {author} {\bibfnamefont {M.}~\bibnamefont {Havey}},\ and\ \bibinfo {author}
  {\bibfnamefont {I.}~\bibnamefont {Sokolov}},\ }\bibfield  {title} {\bibinfo
  {title} {Observation of single-photon superradiance and the cooperative
  {Lamb} shift in an extended sample of cold atoms},\ }\bibfield  {journal}
  {\bibinfo  {journal} {Physical Review Letters}\ }\textbf {\bibinfo {volume}
  {117}},\ \href {https://doi.org/10.1103/physrevlett.117.073003}
  {10.1103/physrevlett.117.073003} (\bibinfo {year} {2016})\BibitemShut
  {NoStop}%
\bibitem [{\citenamefont {Guerin}\ \emph {et~al.}(2016)\citenamefont {Guerin},
  \citenamefont {Ara{\'{u}}jo},\ and\ \citenamefont {Kaiser}}]{Guerin2016}%
  \BibitemOpen
  \bibfield  {author} {\bibinfo {author} {\bibfnamefont {W.}~\bibnamefont
  {Guerin}}, \bibinfo {author} {\bibfnamefont {M.~O.}\ \bibnamefont
  {Ara{\'{u}}jo}},\ and\ \bibinfo {author} {\bibfnamefont {R.}~\bibnamefont
  {Kaiser}},\ }\bibfield  {title} {\bibinfo {title} {Subradiance in a large
  cloud of cold atoms},\ }\bibfield  {journal} {\bibinfo  {journal} {Physical
  Review Letters}\ }\textbf {\bibinfo {volume} {116}},\ \href
  {https://doi.org/10.1103/physrevlett.116.083601}
  {10.1103/physrevlett.116.083601} (\bibinfo {year} {2016})\BibitemShut
  {NoStop}%
\bibitem [{\citenamefont {Ferioli}\ \emph {et~al.}(2021)\citenamefont
  {Ferioli}, \citenamefont {Glicenstein}, \citenamefont {Henriet},
  \citenamefont {Ferrier-Barbut},\ and\ \citenamefont
  {Browaeys}}]{Ferioli2021}%
  \BibitemOpen
  \bibfield  {author} {\bibinfo {author} {\bibfnamefont {G.}~\bibnamefont
  {Ferioli}}, \bibinfo {author} {\bibfnamefont {A.}~\bibnamefont
  {Glicenstein}}, \bibinfo {author} {\bibfnamefont {L.}~\bibnamefont
  {Henriet}}, \bibinfo {author} {\bibfnamefont {I.}~\bibnamefont
  {Ferrier-Barbut}},\ and\ \bibinfo {author} {\bibfnamefont {A.}~\bibnamefont
  {Browaeys}},\ }\bibfield  {title} {\bibinfo {title} {Storage and release of
  subradiant excitations in a dense atomic cloud},\ }\href
  {https://doi.org/10.1103/physrevx.11.021031} {\bibfield  {journal} {\bibinfo
  {journal} {Physical Review X}\ }\textbf {\bibinfo {volume} {11}},\ \bibinfo
  {pages} {021031} (\bibinfo {year} {2021})}\BibitemShut {NoStop}%
\bibitem [{\citenamefont {Goban}\ \emph {et~al.}(2015)\citenamefont {Goban},
  \citenamefont {Hung}, \citenamefont {Hood}, \citenamefont {Yu}, \citenamefont
  {Muniz}, \citenamefont {Painter},\ and\ \citenamefont {Kimble}}]{Goban2015}%
  \BibitemOpen
  \bibfield  {author} {\bibinfo {author} {\bibfnamefont {A.}~\bibnamefont
  {Goban}}, \bibinfo {author} {\bibfnamefont {C.-L.}\ \bibnamefont {Hung}},
  \bibinfo {author} {\bibfnamefont {J.}~\bibnamefont {Hood}}, \bibinfo {author}
  {\bibfnamefont {S.-P.}\ \bibnamefont {Yu}}, \bibinfo {author} {\bibfnamefont
  {J.}~\bibnamefont {Muniz}}, \bibinfo {author} {\bibfnamefont
  {O.}~\bibnamefont {Painter}},\ and\ \bibinfo {author} {\bibfnamefont
  {H.}~\bibnamefont {Kimble}},\ }\bibfield  {title} {\bibinfo {title}
  {Superradiance for atoms trapped along a photonic crystal waveguide},\
  }\bibfield  {journal} {\bibinfo  {journal} {Physical Review Letters}\
  }\textbf {\bibinfo {volume} {115}},\ \href
  {https://doi.org/10.1103/physrevlett.115.063601}
  {10.1103/physrevlett.115.063601} (\bibinfo {year} {2015})\BibitemShut
  {NoStop}%
\bibitem [{\citenamefont {Okaba}\ \emph {et~al.}(2019)\citenamefont {Okaba},
  \citenamefont {Yu}, \citenamefont {Vincetti}, \citenamefont {Benabid},\ and\
  \citenamefont {Katori}}]{Okaba2019}%
  \BibitemOpen
  \bibfield  {author} {\bibinfo {author} {\bibfnamefont {S.}~\bibnamefont
  {Okaba}}, \bibinfo {author} {\bibfnamefont {D.}~\bibnamefont {Yu}}, \bibinfo
  {author} {\bibfnamefont {L.}~\bibnamefont {Vincetti}}, \bibinfo {author}
  {\bibfnamefont {F.}~\bibnamefont {Benabid}},\ and\ \bibinfo {author}
  {\bibfnamefont {H.}~\bibnamefont {Katori}},\ }\bibfield  {title} {\bibinfo
  {title} {Superradiance from lattice-confined atoms inside hollow core
  fibre},\ }\bibfield  {journal} {\bibinfo  {journal} {Communications Physics}\
  }\textbf {\bibinfo {volume} {2}},\ \href
  {https://doi.org/10.1038/s42005-019-0237-2} {10.1038/s42005-019-0237-2}
  (\bibinfo {year} {2019})\BibitemShut {NoStop}%
\bibitem [{\citenamefont {Bettles}\ \emph {et~al.}(2020)\citenamefont
  {Bettles}, \citenamefont {Ilieva}, \citenamefont {Busch}, \citenamefont
  {Huillery}, \citenamefont {Ball}, \citenamefont {Spong},\ and\ \citenamefont
  {Adams}}]{Bettles2020}%
  \BibitemOpen
  \bibfield  {author} {\bibinfo {author} {\bibfnamefont {R.~J.}\ \bibnamefont
  {Bettles}}, \bibinfo {author} {\bibfnamefont {T.}~\bibnamefont {Ilieva}},
  \bibinfo {author} {\bibfnamefont {H.}~\bibnamefont {Busch}}, \bibinfo
  {author} {\bibfnamefont {P.}~\bibnamefont {Huillery}}, \bibinfo {author}
  {\bibfnamefont {S.~W.}\ \bibnamefont {Ball}}, \bibinfo {author}
  {\bibfnamefont {N.~L.~R.}\ \bibnamefont {Spong}},\ and\ \bibinfo {author}
  {\bibfnamefont {C.~S.}\ \bibnamefont {Adams}},\ }\bibfield  {title} {\bibinfo
  {title} {Collective mode interferences in light-matter interactions},\
  }\href@noop {} {\bibfield  {journal} {\bibinfo  {journal}
  {arXiv:1808.08415v4}\ } (\bibinfo {year} {2020})}\BibitemShut {NoStop}%
\bibitem [{\citenamefont {Pennetta}\ \emph {et~al.}(2021)\citenamefont
  {Pennetta}, \citenamefont {Blaha}, \citenamefont {Johnson}, \citenamefont
  {Lechner}, \citenamefont {Schneeweiss}, \citenamefont {Volz},\ and\
  \citenamefont {Rauschenbeutel}}]{Pennetta2021}%
  \BibitemOpen
  \bibfield  {author} {\bibinfo {author} {\bibfnamefont {R.}~\bibnamefont
  {Pennetta}}, \bibinfo {author} {\bibfnamefont {M.}~\bibnamefont {Blaha}},
  \bibinfo {author} {\bibfnamefont {A.}~\bibnamefont {Johnson}}, \bibinfo
  {author} {\bibfnamefont {D.}~\bibnamefont {Lechner}}, \bibinfo {author}
  {\bibfnamefont {P.}~\bibnamefont {Schneeweiss}}, \bibinfo {author}
  {\bibfnamefont {J.}~\bibnamefont {Volz}},\ and\ \bibinfo {author}
  {\bibfnamefont {A.}~\bibnamefont {Rauschenbeutel}},\ }\bibfield  {title}
  {\bibinfo {title} {Collective radiative dynamics of an ensemble of cold atoms
  coupled to an optical waveguide},\ }\href@noop {} {\bibfield  {journal}
  {\bibinfo  {journal} {arXiv:2109.00860}\ } (\bibinfo {year}
  {2021})}\BibitemShut {NoStop}%
\bibitem [{\citenamefont {Paris-Mandoki}\ \emph {et~al.}(2017)\citenamefont
  {Paris-Mandoki}, \citenamefont {Braun}, \citenamefont {Kumlin}, \citenamefont
  {Tresp}, \citenamefont {Mirgorodskiy}, \citenamefont {Christaller},
  \citenamefont {Büchler},\ and\ \citenamefont
  {Hofferberth}}]{ParisMandoki2017}%
  \BibitemOpen
  \bibfield  {author} {\bibinfo {author} {\bibfnamefont {A.}~\bibnamefont
  {Paris-Mandoki}}, \bibinfo {author} {\bibfnamefont {C.}~\bibnamefont
  {Braun}}, \bibinfo {author} {\bibfnamefont {J.}~\bibnamefont {Kumlin}},
  \bibinfo {author} {\bibfnamefont {C.}~\bibnamefont {Tresp}}, \bibinfo
  {author} {\bibfnamefont {I.}~\bibnamefont {Mirgorodskiy}}, \bibinfo {author}
  {\bibfnamefont {F.}~\bibnamefont {Christaller}}, \bibinfo {author}
  {\bibfnamefont {H.~P.}\ \bibnamefont {Büchler}},\ and\ \bibinfo {author}
  {\bibfnamefont {S.}~\bibnamefont {Hofferberth}},\ }\bibfield  {title}
  {\bibinfo {title} {Free-space quantum electrodynamics with a single {Rydberg}
  superatom},\ }\href {https://doi.org/10.1103/physrevx.7.041010} {\bibfield
  {journal} {\bibinfo  {journal} {Physical Review X}\ }\textbf {\bibinfo
  {volume} {7}},\ \bibinfo {pages} {041010} (\bibinfo {year}
  {2017})}\BibitemShut {NoStop}%
\bibitem [{\citenamefont {Stiesdal}\ \emph {et~al.}(2020)\citenamefont
  {Stiesdal}, \citenamefont {Busche}, \citenamefont {Kumlin}, \citenamefont
  {Kleinbeck}, \citenamefont {Büchler},\ and\ \citenamefont
  {Hofferberth}}]{Stiesdal2020}%
  \BibitemOpen
  \bibfield  {author} {\bibinfo {author} {\bibfnamefont {N.}~\bibnamefont
  {Stiesdal}}, \bibinfo {author} {\bibfnamefont {H.}~\bibnamefont {Busche}},
  \bibinfo {author} {\bibfnamefont {J.}~\bibnamefont {Kumlin}}, \bibinfo
  {author} {\bibfnamefont {K.}~\bibnamefont {Kleinbeck}}, \bibinfo {author}
  {\bibfnamefont {H.~P.}\ \bibnamefont {Büchler}},\ and\ \bibinfo {author}
  {\bibfnamefont {S.}~\bibnamefont {Hofferberth}},\ }\bibfield  {title}
  {\bibinfo {title} {Observation of collective decay dynamics of a single
  {Rydberg} superatom},\ }\href
  {https://doi.org/10.1103/physrevresearch.2.043339} {\bibfield  {journal}
  {\bibinfo  {journal} {Physical Review Research}\ }\textbf {\bibinfo {volume}
  {2}},\ \bibinfo {pages} {043339} (\bibinfo {year} {2020})}\BibitemShut
  {NoStop}%
\bibitem [{\citenamefont {Rohlsberger}\ \emph {et~al.}(2010)\citenamefont
  {Rohlsberger}, \citenamefont {Schlage}, \citenamefont {Sahoo}, \citenamefont
  {Couet},\ and\ \citenamefont {Ruffer}}]{Rohlsberger2010}%
  \BibitemOpen
  \bibfield  {author} {\bibinfo {author} {\bibfnamefont {R.}~\bibnamefont
  {Rohlsberger}}, \bibinfo {author} {\bibfnamefont {K.}~\bibnamefont
  {Schlage}}, \bibinfo {author} {\bibfnamefont {B.}~\bibnamefont {Sahoo}},
  \bibinfo {author} {\bibfnamefont {S.}~\bibnamefont {Couet}},\ and\ \bibinfo
  {author} {\bibfnamefont {R.}~\bibnamefont {Ruffer}},\ }\bibfield  {title}
  {\bibinfo {title} {Collective {Lamb} shift in single-photon superradiance},\
  }\href {https://doi.org/10.1126/science.1187770} {\bibfield  {journal}
  {\bibinfo  {journal} {Science}\ }\textbf {\bibinfo {volume} {328}},\ \bibinfo
  {pages} {1248} (\bibinfo {year} {2010})}\BibitemShut {NoStop}%
\bibitem [{\citenamefont {Tavis}\ and\ \citenamefont
  {Cummings}(1968)}]{Tavis1968}%
  \BibitemOpen
  \bibfield  {author} {\bibinfo {author} {\bibfnamefont {M.}~\bibnamefont
  {Tavis}}\ and\ \bibinfo {author} {\bibfnamefont {F.~W.}\ \bibnamefont
  {Cummings}},\ }\bibfield  {title} {\bibinfo {title} {Exact solution for {an
  N}-molecule-radiation-field hamiltonian},\ }\href
  {https://doi.org/10.1103/physrev.170.379} {\bibfield  {journal} {\bibinfo
  {journal} {Physical Review}\ }\textbf {\bibinfo {volume} {170}},\ \bibinfo
  {pages} {379} (\bibinfo {year} {1968})}\BibitemShut {NoStop}%
\bibitem [{\citenamefont {Blaha}\ \emph {et~al.}(2021)\citenamefont {Blaha},
  \citenamefont {Johnson}, \citenamefont {Rauschenbeutel},\ and\ \citenamefont
  {Volz}}]{Blaha2021}%
  \BibitemOpen
  \bibfield  {author} {\bibinfo {author} {\bibfnamefont {M.}~\bibnamefont
  {Blaha}}, \bibinfo {author} {\bibfnamefont {A.}~\bibnamefont {Johnson}},
  \bibinfo {author} {\bibfnamefont {A.}~\bibnamefont {Rauschenbeutel}},\ and\
  \bibinfo {author} {\bibfnamefont {J.}~\bibnamefont {Volz}},\ }\bibfield
  {title} {\bibinfo {title} {Beyond the {Tavis-Cummings} model: revisiting
  cavity {QED} with atomic ensembles},\ }\href@noop {} {\bibfield  {journal}
  {\bibinfo  {journal} {arXiv:2107.04583v1}\ } (\bibinfo {year}
  {2021})}\BibitemShut {NoStop}%
\bibitem [{\citenamefont {Colombe}\ \emph {et~al.}(2007)\citenamefont
  {Colombe}, \citenamefont {Steinmetz}, \citenamefont {Dubois}, \citenamefont
  {Linke}, \citenamefont {Hunger},\ and\ \citenamefont
  {Reichel}}]{Colombe2007}%
  \BibitemOpen
  \bibfield  {author} {\bibinfo {author} {\bibfnamefont {Y.}~\bibnamefont
  {Colombe}}, \bibinfo {author} {\bibfnamefont {T.}~\bibnamefont {Steinmetz}},
  \bibinfo {author} {\bibfnamefont {G.}~\bibnamefont {Dubois}}, \bibinfo
  {author} {\bibfnamefont {F.}~\bibnamefont {Linke}}, \bibinfo {author}
  {\bibfnamefont {D.}~\bibnamefont {Hunger}},\ and\ \bibinfo {author}
  {\bibfnamefont {J.}~\bibnamefont {Reichel}},\ }\bibfield  {title} {\bibinfo
  {title} {Strong atom{\textendash}field coupling for bose{\textendash}einstein
  condensates in an optical cavity on a chip},\ }\href
  {https://doi.org/10.1038/nature06331} {\bibfield  {journal} {\bibinfo
  {journal} {Nature}\ }\textbf {\bibinfo {volume} {450}},\ \bibinfo {pages}
  {272} (\bibinfo {year} {2007})}\BibitemShut {NoStop}%
\bibitem [{\citenamefont {Brennecke}\ \emph {et~al.}(2007)\citenamefont
  {Brennecke}, \citenamefont {Donner}, \citenamefont {Ritter}, \citenamefont
  {Bourdel}, \citenamefont {Köhl},\ and\ \citenamefont
  {Esslinger}}]{Brennecke2007}%
  \BibitemOpen
  \bibfield  {author} {\bibinfo {author} {\bibfnamefont {F.}~\bibnamefont
  {Brennecke}}, \bibinfo {author} {\bibfnamefont {T.}~\bibnamefont {Donner}},
  \bibinfo {author} {\bibfnamefont {S.}~\bibnamefont {Ritter}}, \bibinfo
  {author} {\bibfnamefont {T.}~\bibnamefont {Bourdel}}, \bibinfo {author}
  {\bibfnamefont {M.}~\bibnamefont {Köhl}},\ and\ \bibinfo {author}
  {\bibfnamefont {T.}~\bibnamefont {Esslinger}},\ }\bibfield  {title} {\bibinfo
  {title} {Cavity {QED} with a bose{\textendash}einstein condensate},\ }\href
  {https://doi.org/10.1038/nature06120} {\bibfield  {journal} {\bibinfo
  {journal} {Nature}\ }\textbf {\bibinfo {volume} {450}},\ \bibinfo {pages}
  {268} (\bibinfo {year} {2007})}\BibitemShut {NoStop}%
\bibitem [{\citenamefont {Svidzinsky}\ \emph {et~al.}(2008)\citenamefont
  {Svidzinsky}, \citenamefont {Chang},\ and\ \citenamefont
  {Scully}}]{Svidzinsky2008}%
  \BibitemOpen
  \bibfield  {author} {\bibinfo {author} {\bibfnamefont {A.~A.}\ \bibnamefont
  {Svidzinsky}}, \bibinfo {author} {\bibfnamefont {J.~T.}\ \bibnamefont
  {Chang}},\ and\ \bibinfo {author} {\bibfnamefont {M.~O.}\ \bibnamefont
  {Scully}},\ }\bibfield  {title} {\bibinfo {title} {Dynamical evolution of
  correlated spontaneous emission of a single photon from a uniformly excited
  cloud of {N} atoms},\ }\bibfield  {journal} {\bibinfo  {journal} {Physical
  Review Letters}\ }\textbf {\bibinfo {volume} {100}},\ \href
  {https://doi.org/10.1103/physrevlett.100.160504}
  {10.1103/physrevlett.100.160504} (\bibinfo {year} {2008})\BibitemShut
  {NoStop}%
\bibitem [{\citenamefont {Kumlin}\ \emph {et~al.}(2020)\citenamefont {Kumlin},
  \citenamefont {Kleinbeck}, \citenamefont {Stiesdal}, \citenamefont {Busche},
  \citenamefont {Hofferberth},\ and\ \citenamefont {Büchler}}]{Kumlin2020}%
  \BibitemOpen
  \bibfield  {author} {\bibinfo {author} {\bibfnamefont {J.}~\bibnamefont
  {Kumlin}}, \bibinfo {author} {\bibfnamefont {K.}~\bibnamefont {Kleinbeck}},
  \bibinfo {author} {\bibfnamefont {N.}~\bibnamefont {Stiesdal}}, \bibinfo
  {author} {\bibfnamefont {H.}~\bibnamefont {Busche}}, \bibinfo {author}
  {\bibfnamefont {S.}~\bibnamefont {Hofferberth}},\ and\ \bibinfo {author}
  {\bibfnamefont {H.~P.}\ \bibnamefont {Büchler}},\ }\bibfield  {title}
  {\bibinfo {title} {Nonexponential decay of a collective excitation in an
  atomic ensemble coupled to a one-dimensional waveguide},\ }\bibfield
  {journal} {\bibinfo  {journal} {Physical Review A}\ }\textbf {\bibinfo
  {volume} {102}},\ \href {https://doi.org/10.1103/physreva.102.063703}
  {10.1103/physreva.102.063703} (\bibinfo {year} {2020})\BibitemShut {NoStop}%
\bibitem [{\citenamefont {Jen}\ \emph {et~al.}(2020)\citenamefont {Jen},
  \citenamefont {Chang}, \citenamefont {Lin},\ and\ \citenamefont
  {Chen}}]{Jen2020}%
  \BibitemOpen
  \bibfield  {author} {\bibinfo {author} {\bibfnamefont {H.~H.}\ \bibnamefont
  {Jen}}, \bibinfo {author} {\bibfnamefont {M.-S.}\ \bibnamefont {Chang}},
  \bibinfo {author} {\bibfnamefont {G.-D.}\ \bibnamefont {Lin}},\ and\ \bibinfo
  {author} {\bibfnamefont {Y.-C.}\ \bibnamefont {Chen}},\ }\bibfield  {title}
  {\bibinfo {title} {Subradiance dynamics in a singly excited chirally coupled
  atomic chain},\ }\href {https://doi.org/10.1103/physreva.101.023830}
  {\bibfield  {journal} {\bibinfo  {journal} {Physical Review A}\ }\textbf
  {\bibinfo {volume} {101}},\ \bibinfo {pages} {023830} (\bibinfo {year}
  {2020})}\BibitemShut {NoStop}%
\bibitem [{\citenamefont {Berman}(2020)}]{Berman2020}%
  \BibitemOpen
  \bibfield  {author} {\bibinfo {author} {\bibfnamefont {P.~R.}\ \bibnamefont
  {Berman}},\ }\bibfield  {title} {\bibinfo {title} {Theory of two atoms in a
  chiral waveguide},\ }\href {https://doi.org/10.1103/physreva.101.013830}
  {\bibfield  {journal} {\bibinfo  {journal} {Physical Review A}\ }\textbf
  {\bibinfo {volume} {101}},\ \bibinfo {pages} {013830} (\bibinfo {year}
  {2020})}\BibitemShut {NoStop}%
\bibitem [{\citenamefont {Pivovarov}\ \emph {et~al.}(2021)\citenamefont
  {Pivovarov}, \citenamefont {Gerasimov}, \citenamefont {Berroir},
  \citenamefont {Ray}, \citenamefont {Laurat}, \citenamefont {Urvoy},\ and\
  \citenamefont {Kupriyanov}}]{Pivovarov2021}%
  \BibitemOpen
  \bibfield  {author} {\bibinfo {author} {\bibfnamefont {V.~A.}\ \bibnamefont
  {Pivovarov}}, \bibinfo {author} {\bibfnamefont {L.~V.}\ \bibnamefont
  {Gerasimov}}, \bibinfo {author} {\bibfnamefont {J.}~\bibnamefont {Berroir}},
  \bibinfo {author} {\bibfnamefont {T.}~\bibnamefont {Ray}}, \bibinfo {author}
  {\bibfnamefont {J.}~\bibnamefont {Laurat}}, \bibinfo {author} {\bibfnamefont
  {A.}~\bibnamefont {Urvoy}},\ and\ \bibinfo {author} {\bibfnamefont {D.~V.}\
  \bibnamefont {Kupriyanov}},\ }\bibfield  {title} {\bibinfo {title} {Single
  collective excitation of an atomic array trapped along a waveguide: A study
  of cooperative emission for different atomic chain configurations},\ }\href
  {https://doi.org/10.1103/physreva.103.043716} {\bibfield  {journal} {\bibinfo
   {journal} {Physical Review A}\ }\textbf {\bibinfo {volume} {103}},\ \bibinfo
  {pages} {043716} (\bibinfo {year} {2021})}\BibitemShut {NoStop}%
\bibitem [{\citenamefont {Kien}\ and\ \citenamefont
  {Rauschenbeutel}(2017)}]{Kien2017}%
  \BibitemOpen
  \bibfield  {author} {\bibinfo {author} {\bibfnamefont {F.~L.}\ \bibnamefont
  {Kien}}\ and\ \bibinfo {author} {\bibfnamefont {A.}~\bibnamefont
  {Rauschenbeutel}},\ }\bibfield  {title} {\bibinfo {title} {Nanofiber-mediated
  chiral radiative coupling between two atoms},\ }\href
  {https://doi.org/10.1103/physreva.95.023838} {\bibfield  {journal} {\bibinfo
  {journal} {Physical Review A}\ }\textbf {\bibinfo {volume} {95}},\ \bibinfo
  {pages} {023838} (\bibinfo {year} {2017})}\BibitemShut {NoStop}%
\bibitem [{\citenamefont {Vetsch}\ \emph {et~al.}(2010)\citenamefont {Vetsch},
  \citenamefont {Reitz}, \citenamefont {Sagu{\'{e}}}, \citenamefont {Schmidt},
  \citenamefont {Dawkins},\ and\ \citenamefont {Rauschenbeutel}}]{Vetsch2010}%
  \BibitemOpen
  \bibfield  {author} {\bibinfo {author} {\bibfnamefont {E.}~\bibnamefont
  {Vetsch}}, \bibinfo {author} {\bibfnamefont {D.}~\bibnamefont {Reitz}},
  \bibinfo {author} {\bibfnamefont {G.}~\bibnamefont {Sagu{\'{e}}}}, \bibinfo
  {author} {\bibfnamefont {R.}~\bibnamefont {Schmidt}}, \bibinfo {author}
  {\bibfnamefont {S.~T.}\ \bibnamefont {Dawkins}},\ and\ \bibinfo {author}
  {\bibfnamefont {A.}~\bibnamefont {Rauschenbeutel}},\ }\bibfield  {title}
  {\bibinfo {title} {Optical interface created by laser-cooled atoms trapped in
  the evanescent field surrounding an optical nanofiber},\ }\bibfield
  {journal} {\bibinfo  {journal} {Physical Review Letters}\ }\textbf {\bibinfo
  {volume} {104}},\ \href {https://doi.org/10.1103/physrevlett.104.203603}
  {10.1103/physrevlett.104.203603} (\bibinfo {year} {2010})\BibitemShut
  {NoStop}%
\bibitem [{\citenamefont {Thompson}\ \emph {et~al.}(2013)\citenamefont
  {Thompson}, \citenamefont {Tiecke}, \citenamefont {de~Leon}, \citenamefont
  {Feist}, \citenamefont {Akimov}, \citenamefont {Gullans}, \citenamefont
  {Zibrov}, \citenamefont {Vuletic},\ and\ \citenamefont
  {Lukin}}]{Thompson2013}%
  \BibitemOpen
  \bibfield  {author} {\bibinfo {author} {\bibfnamefont {J.~D.}\ \bibnamefont
  {Thompson}}, \bibinfo {author} {\bibfnamefont {T.~G.}\ \bibnamefont
  {Tiecke}}, \bibinfo {author} {\bibfnamefont {N.~P.}\ \bibnamefont {de~Leon}},
  \bibinfo {author} {\bibfnamefont {J.}~\bibnamefont {Feist}}, \bibinfo
  {author} {\bibfnamefont {A.~V.}\ \bibnamefont {Akimov}}, \bibinfo {author}
  {\bibfnamefont {M.}~\bibnamefont {Gullans}}, \bibinfo {author} {\bibfnamefont
  {A.~S.}\ \bibnamefont {Zibrov}}, \bibinfo {author} {\bibfnamefont
  {V.}~\bibnamefont {Vuletic}},\ and\ \bibinfo {author} {\bibfnamefont {M.~D.}\
  \bibnamefont {Lukin}},\ }\bibfield  {title} {\bibinfo {title} {Coupling a
  single trapped atom to a nanoscale optical cavity},\ }\href
  {https://doi.org/10.1126/science.1237125} {\bibfield  {journal} {\bibinfo
  {journal} {Science}\ }\textbf {\bibinfo {volume} {340}},\ \bibinfo {pages}
  {1202} (\bibinfo {year} {2013})}\BibitemShut {NoStop}%
\bibitem [{\citenamefont {Shen}\ and\ \citenamefont {Fan}(2009)}]{Shen2009}%
  \BibitemOpen
  \bibfield  {author} {\bibinfo {author} {\bibfnamefont {J.-T.}\ \bibnamefont
  {Shen}}\ and\ \bibinfo {author} {\bibfnamefont {S.}~\bibnamefont {Fan}},\
  }\bibfield  {title} {\bibinfo {title} {Theory of single-photon transport in a
  single-mode waveguide. i. coupling to a cavity containing a two-level atom},\
  }\bibfield  {journal} {\bibinfo  {journal} {Physical Review A}\ }\textbf
  {\bibinfo {volume} {79}},\ \href {https://doi.org/10.1103/physreva.79.023837}
  {10.1103/physreva.79.023837} (\bibinfo {year} {2009})\BibitemShut {NoStop}%
\bibitem [{\citenamefont {van Bürck}(1999)}]{Buerck1999}%
  \BibitemOpen
  \bibfield  {author} {\bibinfo {author} {\bibfnamefont {U.}~\bibnamefont {van
  Bürck}},\ }\bibfield  {title} {\bibinfo {title} {Coherent pulse propagation
  through resonant media},\ }\href {https://doi.org/10.1023/a:1017080008712}
  {\bibfield  {journal} {\bibinfo  {journal} {Hyperfine Interactions}\ }\textbf
  {\bibinfo {volume} {123/124}},\ \bibinfo {pages} {483} (\bibinfo {year}
  {1999})}\BibitemShut {NoStop}%
\bibitem [{\citenamefont {Steck}(2019)}]{Steck2019}%
  \BibitemOpen
  \bibfield  {author} {\bibinfo {author} {\bibfnamefont {D.}~\bibnamefont
  {Steck}},\ }\bibfield  {title} {\bibinfo {title} {Cesium d line data},\
  }\href@noop {} {\bibfield  {journal} {\bibinfo  {journal} {available online
  at http://steck.us/alkalidata}\ } (\bibinfo {year} {2019})}\BibitemShut
  {NoStop}%
\bibitem [{\citenamefont {Johnson}\ \emph {et~al.}(2019)\citenamefont
  {Johnson}, \citenamefont {Blaha}, \citenamefont {Ulanov}, \citenamefont
  {Rauschenbeutel}, \citenamefont {Schneeweiss},\ and\ \citenamefont
  {Volz}}]{Johnson2019}%
  \BibitemOpen
  \bibfield  {author} {\bibinfo {author} {\bibfnamefont {A.}~\bibnamefont
  {Johnson}}, \bibinfo {author} {\bibfnamefont {M.}~\bibnamefont {Blaha}},
  \bibinfo {author} {\bibfnamefont {A.~E.}\ \bibnamefont {Ulanov}}, \bibinfo
  {author} {\bibfnamefont {A.}~\bibnamefont {Rauschenbeutel}}, \bibinfo
  {author} {\bibfnamefont {P.}~\bibnamefont {Schneeweiss}},\ and\ \bibinfo
  {author} {\bibfnamefont {J.}~\bibnamefont {Volz}},\ }\bibfield  {title}
  {\bibinfo {title} {Observation of collective superstrong coupling of cold
  atoms to a 30-m long optical resonator},\ }\bibfield  {journal} {\bibinfo
  {journal} {Physical Review Letters}\ }\textbf {\bibinfo {volume} {123}},\
  \href {https://doi.org/10.1103/physrevlett.123.243602}
  {10.1103/physrevlett.123.243602} (\bibinfo {year} {2019})\BibitemShut
  {NoStop}%
\bibitem [{\citenamefont {Kien}\ \emph {et~al.}(2005)\citenamefont {Kien},
  \citenamefont {Gupta}, \citenamefont {Balykin},\ and\ \citenamefont
  {Hakuta}}]{Kien2005}%
  \BibitemOpen
  \bibfield  {author} {\bibinfo {author} {\bibfnamefont {F.~L.}\ \bibnamefont
  {Kien}}, \bibinfo {author} {\bibfnamefont {S.~D.}\ \bibnamefont {Gupta}},
  \bibinfo {author} {\bibfnamefont {V.~I.}\ \bibnamefont {Balykin}},\ and\
  \bibinfo {author} {\bibfnamefont {K.}~\bibnamefont {Hakuta}},\ }\bibfield
  {title} {\bibinfo {title} {Spontaneous emission of a cesium atom near a
  nanofiber: Efficient coupling of light to guided modes},\ }\bibfield
  {journal} {\bibinfo  {journal} {Physical Review A}\ }\textbf {\bibinfo
  {volume} {72}},\ \href {https://doi.org/10.1103/physreva.72.032509}
  {10.1103/physreva.72.032509} (\bibinfo {year} {2005})\BibitemShut {NoStop}%
\bibitem [{\citenamefont {Lvovsky}\ \emph {et~al.}(2009)\citenamefont
  {Lvovsky}, \citenamefont {Sanders},\ and\ \citenamefont
  {Tittel}}]{Lvovsky2009}%
  \BibitemOpen
  \bibfield  {author} {\bibinfo {author} {\bibfnamefont {A.~I.}\ \bibnamefont
  {Lvovsky}}, \bibinfo {author} {\bibfnamefont {B.~C.}\ \bibnamefont
  {Sanders}},\ and\ \bibinfo {author} {\bibfnamefont {W.}~\bibnamefont
  {Tittel}},\ }\bibfield  {title} {\bibinfo {title} {Optical quantum memory},\
  }\href {https://doi.org/10.1038/nphoton.2009.231} {\bibfield  {journal}
  {\bibinfo  {journal} {Nature Photonics}\ }\textbf {\bibinfo {volume} {3}},\
  \bibinfo {pages} {706} (\bibinfo {year} {2009})}\BibitemShut {NoStop}%
\bibitem [{\citenamefont {Simon}\ \emph {et~al.}(2010)\citenamefont {Simon},
  \citenamefont {Afzelius}, \citenamefont {Appel}, \citenamefont {de~la
  Giroday}, \citenamefont {Dewhurst}, \citenamefont {Gisin}, \citenamefont
  {Hu}, \citenamefont {Jelezko}, \citenamefont {Kröll}, \citenamefont
  {Müller}, \citenamefont {Nunn}, \citenamefont {Polzik}, \citenamefont
  {Rarity}, \citenamefont {Riedmatten}, \citenamefont {Rosenfeld},
  \citenamefont {Shields}, \citenamefont {Sköld}, \citenamefont {Stevenson},
  \citenamefont {Thew}, \citenamefont {Walmsley}, \citenamefont {Weber},
  \citenamefont {Weinfurter}, \citenamefont {Wrachtrup},\ and\ \citenamefont
  {Young}}]{Simon2010}%
  \BibitemOpen
  \bibfield  {author} {\bibinfo {author} {\bibfnamefont {C.}~\bibnamefont
  {Simon}}, \bibinfo {author} {\bibfnamefont {M.}~\bibnamefont {Afzelius}},
  \bibinfo {author} {\bibfnamefont {J.}~\bibnamefont {Appel}}, \bibinfo
  {author} {\bibfnamefont {A.~B.}\ \bibnamefont {de~la Giroday}}, \bibinfo
  {author} {\bibfnamefont {S.~J.}\ \bibnamefont {Dewhurst}}, \bibinfo {author}
  {\bibfnamefont {N.}~\bibnamefont {Gisin}}, \bibinfo {author} {\bibfnamefont
  {C.~Y.}\ \bibnamefont {Hu}}, \bibinfo {author} {\bibfnamefont
  {F.}~\bibnamefont {Jelezko}}, \bibinfo {author} {\bibfnamefont
  {S.}~\bibnamefont {Kröll}}, \bibinfo {author} {\bibfnamefont {J.~H.}\
  \bibnamefont {Müller}}, \bibinfo {author} {\bibfnamefont {J.}~\bibnamefont
  {Nunn}}, \bibinfo {author} {\bibfnamefont {E.~S.}\ \bibnamefont {Polzik}},
  \bibinfo {author} {\bibfnamefont {J.~G.}\ \bibnamefont {Rarity}}, \bibinfo
  {author} {\bibfnamefont {H.~D.}\ \bibnamefont {Riedmatten}}, \bibinfo
  {author} {\bibfnamefont {W.}~\bibnamefont {Rosenfeld}}, \bibinfo {author}
  {\bibfnamefont {A.~J.}\ \bibnamefont {Shields}}, \bibinfo {author}
  {\bibfnamefont {N.}~\bibnamefont {Sköld}}, \bibinfo {author} {\bibfnamefont
  {R.~M.}\ \bibnamefont {Stevenson}}, \bibinfo {author} {\bibfnamefont
  {R.}~\bibnamefont {Thew}}, \bibinfo {author} {\bibfnamefont {I.~A.}\
  \bibnamefont {Walmsley}}, \bibinfo {author} {\bibfnamefont {M.~C.}\
  \bibnamefont {Weber}}, \bibinfo {author} {\bibfnamefont {H.}~\bibnamefont
  {Weinfurter}}, \bibinfo {author} {\bibfnamefont {J.}~\bibnamefont
  {Wrachtrup}},\ and\ \bibinfo {author} {\bibfnamefont {R.~J.}\ \bibnamefont
  {Young}},\ }\bibfield  {title} {\bibinfo {title} {Quantum memories},\ }\href
  {https://doi.org/10.1140/epjd/e2010-00103-y} {\bibfield  {journal} {\bibinfo
  {journal} {The European Physical Journal D}\ }\textbf {\bibinfo {volume}
  {58}},\ \bibinfo {pages} {1} (\bibinfo {year} {2010})}\BibitemShut {NoStop}%
\bibitem [{\citenamefont {Mivehvar}\ \emph {et~al.}(2021)\citenamefont
  {Mivehvar}, \citenamefont {Piazza}, \citenamefont {Donner},\ and\
  \citenamefont {Ritsch}}]{Mivehvar2021}%
  \BibitemOpen
  \bibfield  {author} {\bibinfo {author} {\bibfnamefont {F.}~\bibnamefont
  {Mivehvar}}, \bibinfo {author} {\bibfnamefont {F.}~\bibnamefont {Piazza}},
  \bibinfo {author} {\bibfnamefont {T.}~\bibnamefont {Donner}},\ and\ \bibinfo
  {author} {\bibfnamefont {H.}~\bibnamefont {Ritsch}},\ }\bibfield  {title}
  {\bibinfo {title} {Cavity {QED} with quantum gases: new paradigms in
  many-body physics},\ }\href {https://doi.org/10.1080/00018732.2021.1969727}
  {\bibfield  {journal} {\bibinfo  {journal} {Advances in Physics}\ }\textbf
  {\bibinfo {volume} {70}},\ \bibinfo {pages} {1} (\bibinfo {year}
  {2021})}\BibitemShut {NoStop}%
\end{thebibliography}%


\providecommand{\noopsort}[1]{}\providecommand{\singleletter}[1]{#1}%
%

\newpage
\appendix

\section{Supplemental Material}

\begin{figure}[b]
\includegraphics[width=0.95\linewidth]{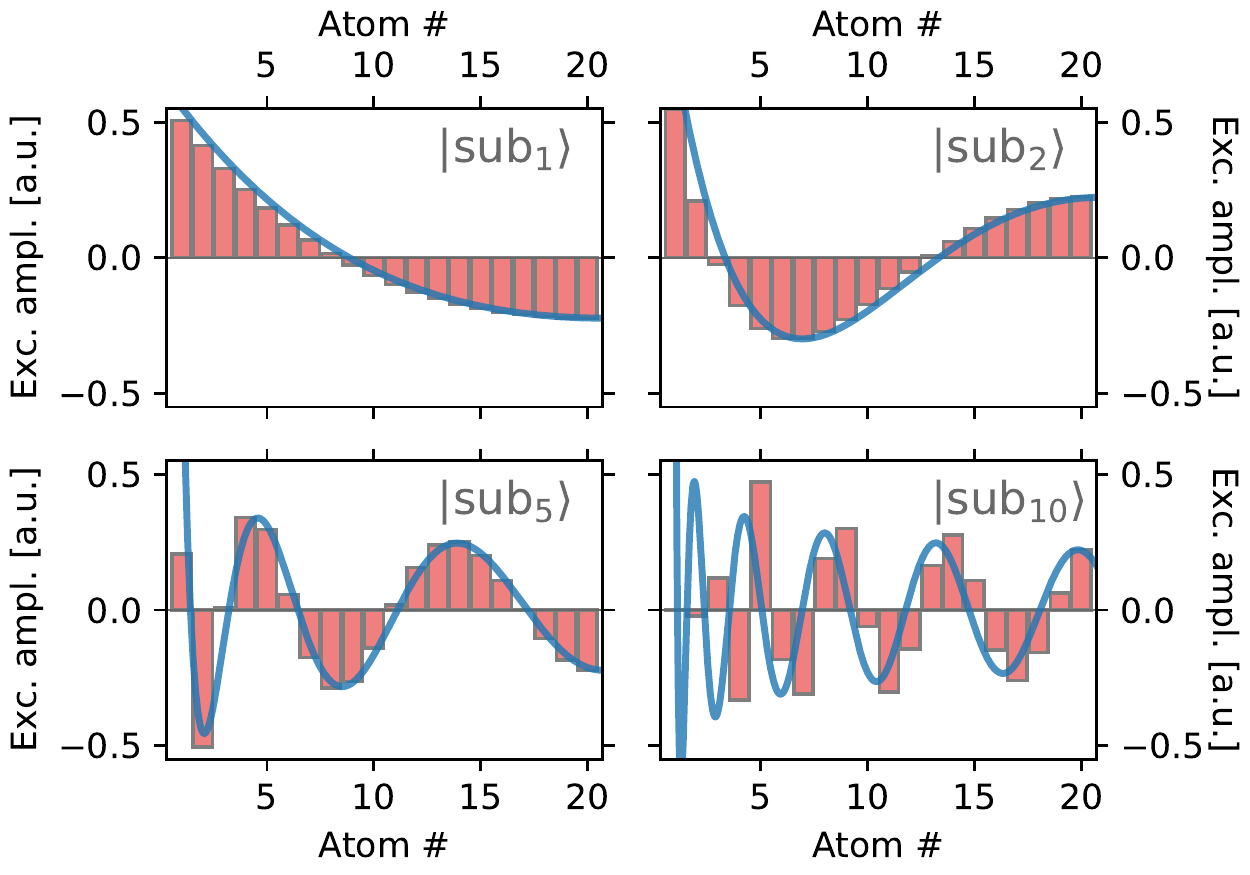}
\caption{\label{fig:ApproxSolutions} Calculation via the exact solution (bars) and the Bessel function approximation (solid line) of the excited state population amplitudes of $N=20$ atoms for the $1^{st}$, $2^{nd}$, $5^{th}$ and $10^{th}$ subradiant states.
}
\end{figure}

\begin{figure}[]
\includegraphics[width=0.92\linewidth]{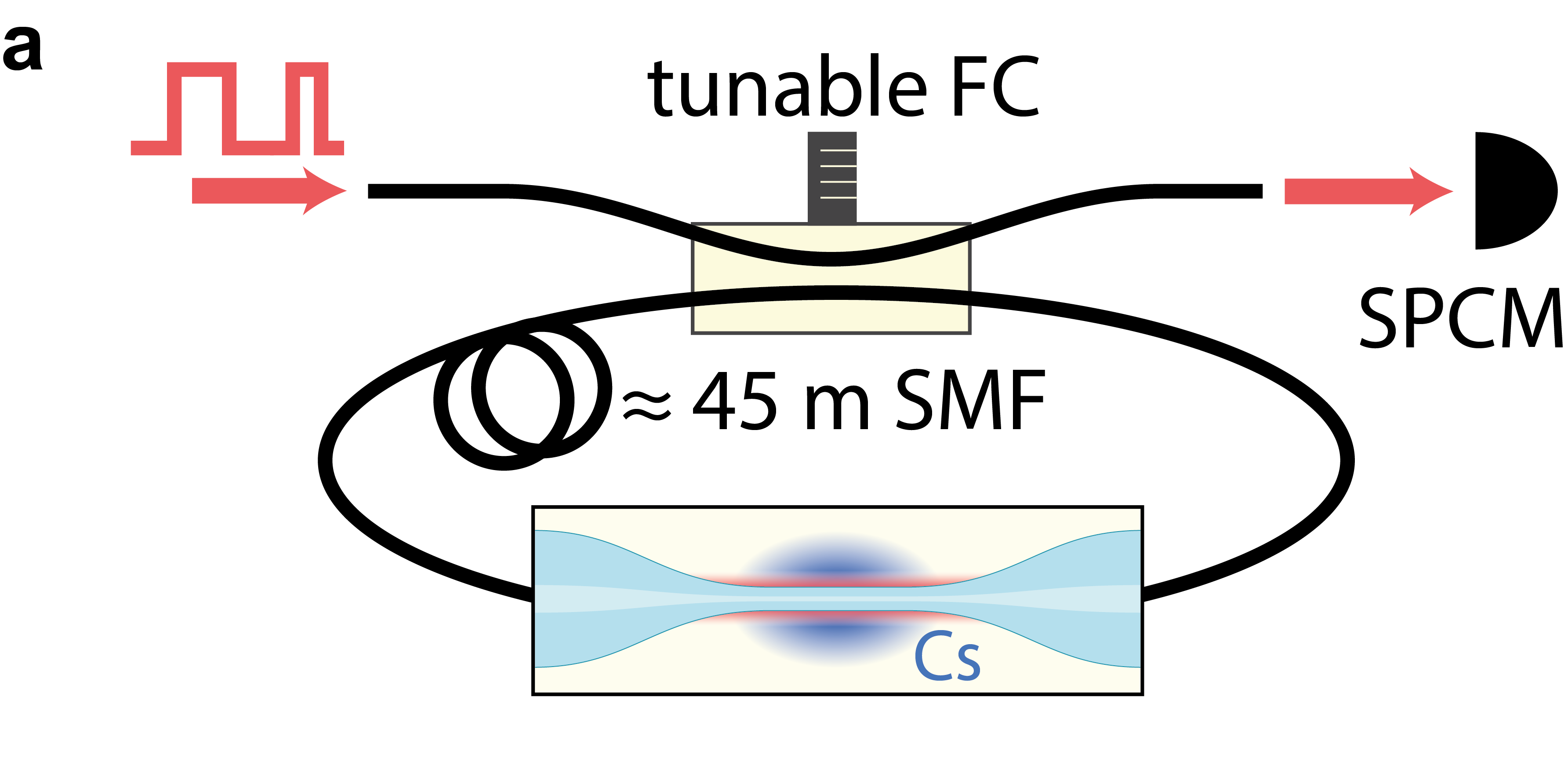}
\includegraphics[width=0.96\linewidth]{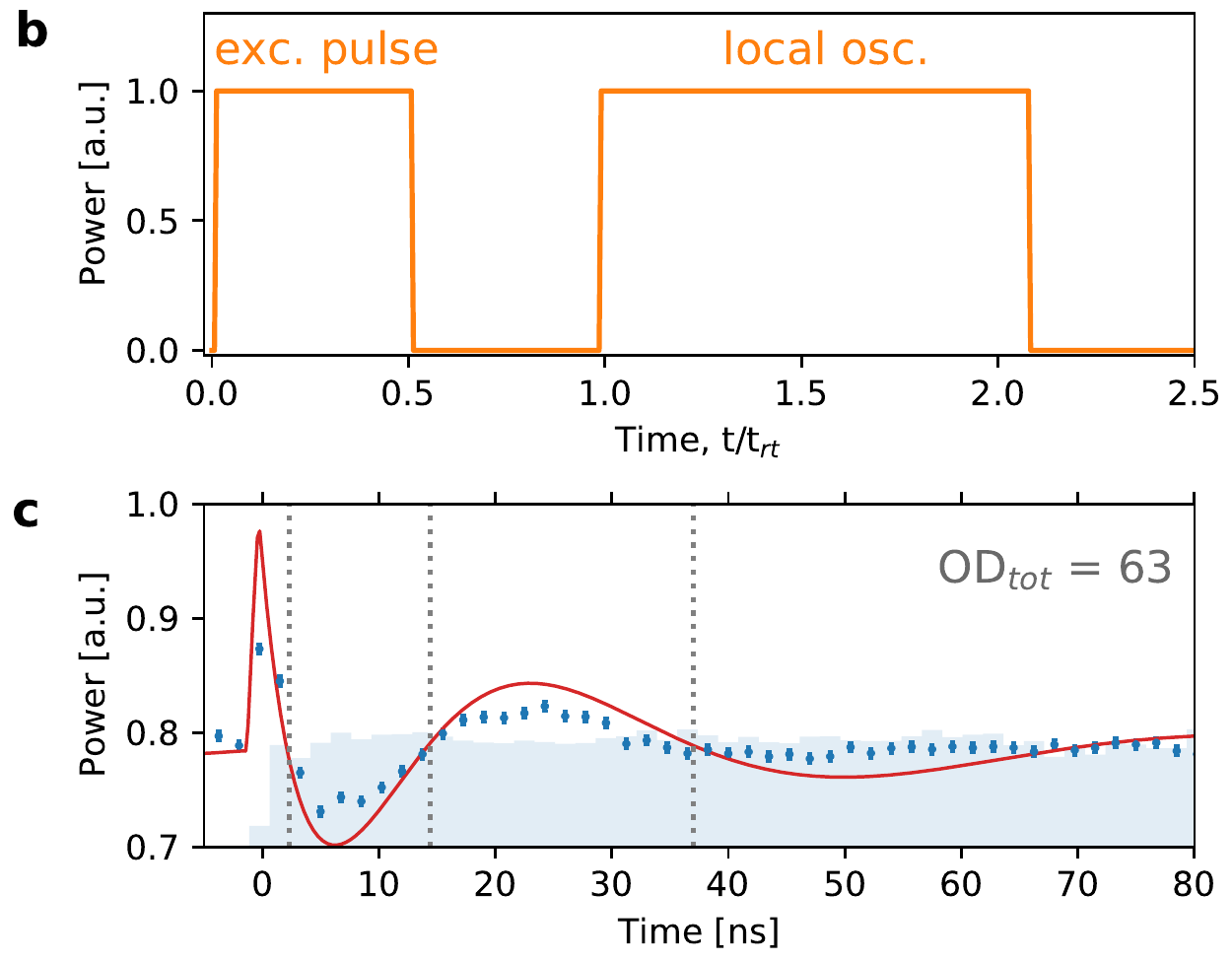}
\caption{\label{fig:LocalOscillator} Setup (a) and pulse sequence (b) used to measure the interference between the atom emitted light and a local oscillator shown in Fig.~\ref{fig:Exp_data}(c) (FC: fiber coupler, SPCM: single-photon counting module, SMF: single-mode fiber, Cs: Cesium, t$_{rt}$: roundtrip time). (c) Measured interference between a local oscillator and the light out-coupled from the ring-resonator after $m$=3 roundtrip. The blue dots are experimental data, while the red solid line indicated out theoretical predictions. The dotted dashed vertical lines indicate the moments in time in which the state of the ensemble passes through the first three subradiant states.
}
\end{figure}

\subsection{Details about the theoretical model}

To theoretically describe light-matter interaction in an ensemble of waveguide-coupled atoms, we follow the formalism of Refs. \cite{Shen2009, Blaha2021, Pennetta2021}. In brief, we start by establishing a real-space non-Hermitian Hamiltonian for $N$ two-level atoms coupled to a single optical mode:
\begin{eqnarray}
\dfrac{H}{\hbar}= \int_{-\infty}^{\infty} \bigg[ a_z^{\dagger} (-i v_g \dfrac{\partial}{\partial z}) a_z +
\sum_{n=1}^N \delta(z-z_n) \Omega_{a,n}  \sigma_n^+\sigma_n^- +
\nonumber\\*
\sum_{n=1}^N \delta(z-z_n)V_n(\sigma_n^+a_z+\sigma_n^-a_z^{\dagger})
\bigg] dz ~,
\qquad \qquad 
\label{eq:Hamiltonian}
\end{eqnarray}
where $a_z^{\dagger}$ ($a_z$) is the creation (annihilation) operator of a photon at position $z$, $\sigma_n^+$ ($\sigma_n^-$) is the raising (lowering) operator for the $n^{\textrm{th}}$ atom at position $z_n$, and $v_g$ is the group velocity of the waveguide mode. Moreover, $\Omega_{a,n} = (\omega_a-i(1-\beta_n) \gamma)$ and $V_n = \sqrt{v_g \beta_n \Gamma_0}$, where $\omega_a$ is the atomic resonance frequency.
We then proceed by calculating the amplitude transmission for the waveguide light in the single-excitation regime for an ensemble of $N$ atoms:
\begin{eqnarray}
t_N( \Delta )= \prod_{n=1}^{N} t_{\textrm{at},n} = \left( 1 - \dfrac{2 \beta_n \gamma}{\gamma + i \Delta} \right)^N ~,
\label{eq:Transmission}
\end{eqnarray}
where $t_{\textrm{at},n}$ is the amplitude transmission of the $n^\textrm{th}$ atom, $\Delta=\omega - \omega_a$ is the laser-atom detuning. In a similar fashion, we can also calculate the probability amplitude of finding the $n^\textrm{th}$ atom in the excited state, $\phi_{n}( \Delta )$: 
\begin{eqnarray}
\phi_{n}( \Delta )=i \dfrac{\sqrt{v_g}}{\sqrt{\beta_n \gamma}}(t_n(\Delta)-t_{n-1}(\Delta)) ~.
\label{eq:ExcProb}
\end{eqnarray}
In the low excitation regime, the transmitted optical field in the time-domain after excitation with a pulse with scalar field amplitude $u_{in}(t)$ can be calculated as:
\begin{eqnarray}
u_{out}(t) = \mathcal{F}^{-1}[{u_{in}(\omega) \cdot t_N(\omega)]} ~,
\label{eq:OutputField}
\end{eqnarray}
where  $\mathcal{F}^{-1}$ indicates the inverse Fourier transform. The time evolution of the excitation amplitude for each of the $N$ atoms follows from an similar derivation.

\subsection{Ensemble decay rate and pulse decay rate for the timed Dicke state}

The ensemble decay rate, $\Gamma_{ens}(t)$, and the decay rate of the atom-emitted light intensity, $\Gamma_{light}(t)$, have been defined in Eqs.~(\ref{eq:Gamma_coll}) and (\ref{eq:Gamma_light}) of the main text, respectively. Their value for an ensemble prepared in the timed Dicke state at $t=0$ can be analytically calculated using the following property of the Laguerre polynomials:
\begin{equation}
\dfrac{d}{dt}L_n^{(0)}(t)=\dfrac{d}{dt}L_{n-1}^{(0)}(t) - L_{n-1}^{(0)}(t) ~,
\end{equation}
yielding the result:
\begin{equation}
\Gamma_{ens}(t=0) = \Gamma_{light}(t=0) = N \beta 2 \gamma + 2\gamma(1-\beta) ~.
\end{equation}
We note that this relation holds only for an ensemble in the timed Dicke state and, therefore, inferring $\Gamma_{ens}(t)$ from the measurement of $\Gamma_{light}(t)$ for any other ensemble state leads to wrong results.

\subsection{Solutions for Heaviside $\theta$-like excitation pulses}

The timed Dicke state can be prepared by exciting the ensemble with Dirac $\delta$-like excitation pulses: $u_{in}(t)~\propto~\delta(t)$. Our model allows to express the resulting $\phi_n(t)$ and $\Phi _n(t)$ in terms of Laguerre polynomials, see Eqs.~(\ref{eq:SolutionsTimedDickeStateAtoms}) and (\ref{eq:SolutionsTimedDickeStateLight}). In our experiment we excite the ensemble with long boxcar-shaped pulse, which prepare a state whose time dynamics does not differ qualitatively from the one of the timed Dicke state. The theoretical predictions depicted in Fig.~\ref{fig:Exp_data} have been calculated by solving numerically Eqs.~(\ref{eq:ExcProb}) and (\ref{eq:OutputField}). However, if we approximate our excitation pulses with Heaviside $\theta$-like pulses (i.e., $u_{in}(t) \propto \theta(-t)$), it is still possible to derive analytic solutions in terms of Laguerre polynmials plus  additional terms. For instance, the calculated solutions for the excites state populations for the case $N=4$ can be expressed as:
\begin{align}
\phi_1(t) =& \dfrac{e^{-\gamma t}}{\sqrt{N}} L_{0}^{(0)}(2 \beta \gamma t) ~,
\\
\phi_2(t) =& \dfrac{e^{-\gamma t}}{\sqrt{N}} \left[ L_{1}^{(0)}(2 \beta \gamma t) - 2\beta \right]  ~,
\\
\phi_3(t) =& \dfrac{e^{-\gamma t}}{\sqrt{N}} \left[ L_{2}^{(0)}(2 \beta \gamma t) - 4\beta 4 \beta^2 (1+\gamma t) \right] ~,
\\
\phi_4(t) =& \dfrac{e^{-\gamma t}}{\sqrt{N}} \Big[ L_{3}^{(0)}(2 \beta \gamma t) - 2\beta(3+6 \beta - 4 \beta^2) +
\\
+& 4 \beta^2(3t\gamma -2 \beta t \gamma - \beta^2 t^2 \gamma^2) \Big] ~. \nonumber
\label{eq:SolutionExpPulses}
\end{align}
These solutions have the same properties as the expressions in Eq. (\ref{eq:SolutionsTimedDickeStateAtoms}), i.e., for the $n^\textrm{th}$ atom they are polynomials with $n-1$ distinct zeros. Therefore, the time dynamics of the system is qualitatively unchanged. However, the exact shape of the $N-1$ subradiant states and the moments in time in which the system reaches them is different.

\subsection{Approximation for $N \gg 1$ and $\beta \ll 1$}

Eqs.~(\ref{eq:SolutionsTimedDickeStateAtomsBessel}) and (\ref{eq:SolutionsTimedDickeStateLightBessel}) of the main text have been obtained by making use of the fact that for large $n$ and fixed $\alpha$ the Laguerre polynomial $L_n^{(\alpha)}(x)$ can be approximated as:
\begin{eqnarray}
L_n^{(\alpha)}(x) \approx \frac{n^\alpha}{\sqrt{x}^\alpha} e^{x/2} J_\alpha(2\sqrt{nx}) ~,
\end{eqnarray}
where $J_\alpha(x)$ are the Bessel functions of the first kind. To illustrate the limit of validity of the approximation via Bessel functions, we compare in Fig.~\ref{fig:ApproxSolutions} the $1^{st}$, $2^{nd}$, $5^{th}$ and $10^{th}$ subradiant states for $N=20$ calculated via the approximate (solid lines) and exact (bars) solutions. Even for relatively small atom number, the Bessel functions approximation shows little deviation from the exact solution as long as $n \ll N$. For higher order subradiant states the two solutions show larger deviations. This is to be expected considering that the Bessel function approximation predicts an infinite number of subradiant states, while the exact solution predicts only $N-1$ of them.

\subsection{Orthogonality of the subradiant states} 
In the main text, we mentioned that the subradiant states of Eq. (\ref{eq:SubradiantStates}) are all orthogonal to each other. While at the current stage we cannot provide a general demonstration based on Eqs.~(\ref{eq:SolutionsTimedDickeStateAtoms}) and (\ref{eq:SolutionsTimedDickeStateLight}), we have checked numerically that this holds true up to $N=200$. In addition, for $N \gg 1$ and small $\beta$, this can be shown using the properties of the Bessel functions. From Eqs.~(\ref{eq:SolutionsTimedDickeStateAtomsBessel}) and (\ref{eq:SolutionsTimedDickeStateLightBessel}) the orthogonality of the $i^\textrm{th}$ and $j^\textrm{th}$ subradiant states (with $i \neq j$) can be demonstrated by showing that:
\begin{eqnarray}
\int_{0}^NJ_0(x_{i}\sqrt{\tfrac{n-1}{N}}) J_0(x_{j}\sqrt{\tfrac{n-1}{N}}) dn = 0 \, ,
\label{eq:Overlap}
\end{eqnarray}
where $x_m$ is the $m^\textrm{th}$ zero of the Bessel function $J_1(x)$. Using the substitution $(n-1)/N=\xi^2$ and $dn=2N \xi d\xi$ we can rewrite the left-hand side of Eq. (\ref{eq:Overlap}) as: 
\begin{eqnarray*}
& & 2N\int_{0}^1 \xi J_0(x_{i}\xi) J_0(x_{j}\xi) d\xi = \\ 
& =&\dfrac{x_{i} J_0(x_{j}) J_1(x_{i}) -x_{j} J_0(x_{i})J_1(x_{j})}{x_{i}^2-x_{j}^2} \underbrace{=}_{i\neq j} 0
\end{eqnarray*}
which proves the result.


\subsection{Experimental sequence}
The experiments, whose results are shown in Fig.~\ref{fig:Exp_data}, are performed by alternating a preparatory phase, in which Cs atoms are loaded into the MOT, and a probing phase, during which the MOT is released for 0.2 ms and 20 probe pulses are launched into the nanofiber. The MOT is then switched-on again for 200 ms to recapture and cool the Cs atoms. The experimental data shown in Fig.~\ref{fig:Exp_data} has been obtained by averaging over $\approx 10^5$ single experimental realization.

~\\

\subsection{Interference with local oscillator}
To experimentally measure the interference between the atom emitted light and a local oscillator, we insert the nanofiber into a $\approx$ 45-m long optical resonator with a roundtrip time $t_{rt}$=219.7~ns, see Fig.~\ref{fig:LocalOscillator}(a). We excite the atoms with a boxcar shaped pulse of duration $t_{exc.}$=110~ns, which is coupled into the ring resonator via a tunable fiber coupler (FC). After a single roundtrip in the cavity, the light transmitted through the atoms is partially out-coupled at the FC and interfered with a second boxcar shaped pulse of duration $t_{LO}=240$~ns, i.e., the local oscillator. Fig.~\ref{fig:LocalOscillator}(b) illustrates the experimental pulse sequence in more detail. We note that light emitted by the atomic ensemble exhibits a phase shift of $\pi$ with respect to the excitation pulse, which itself had already accumulated an additional $\pi$ phase shift due to the coupling to the ring resonator. As a consequence, constructive interference with the local oscillator is expected at $t=0$, in agreement with our measurement in Fig.~\ref{fig:Exp_data}(c). 

We have recently shown that the results obtained by measuring the light out-coupled from the ring-resonator after the $m^\textrm{th}$ roundtrip are equivalent to a single propagation through an ensemble with optical depth OD$_{tot}=m \cdot $ OD$_{sp}$, where OD$_{sp}$ is the single-pass OD \cite{Pennetta2021}. This allows us to perform experiments with large optical depths, which cannot be easily achieved in a single-pass configuration. As an example, we show in Fig.~\ref{fig:LocalOscillator}(c) the results obtained for OD$_{tot}= 3 \cdot$OD$_{sp}$ = 63, in which the passage of the state of the ensemble through the first three subradiant states could be clearly observed.

\end{document}